%% file: main.tex
\documentclass[sigconf, nonacm]{acmart}

\input{datalabmacros.tex}

\begin{document}

\input{title}

\input{abstract}

\maketitle

\input{introduction}

\input{learnedfilters}

\input{stackedfilters}

\input{adaptivefilters}
\input{experiments}
\input{analysis}

\input{discussion}

\iflongversion \input{ack} \fi

\clearpage
\balance
\bibliographystyle{ACM-Reference-Format}
\bibliography{bibliography_filters}

\iflongversion
\clearpage
\nobalance
\appendix
\nobalance
\input{appendix}
\input{filters}

\input{app-learned}
\input{app-stacked}
\input{app-adaptive}
\input{questions}

\input{app-experiments}
\fi

\end{document}

%% file: title.tex
\newcommand{\ourTitle}{How to Train Your Filter: Should You Learn, Stack or Adapt?}
\title[How to Train Your Filter: Should You Learn, Stack or Adapt?]{\ourTitle}

\author{Diandre Miguel Sabale}
\orcid{0009-0005-7689-1756}%
\affiliation{%
    \orcidicon{0009-0005-7689-1756}
	\institution{Northeastern University}
        \city{Boston}
        \state{MA}
        \country{USA}}
\email{sabale.d@northeastern.edu}

\author{Wolfgang Gatterbauer}
\orcid{0000-0002-9614-0504}%
\affiliation{%
    \orcidicon{0000-0002-9614-0504}
	\institution{Northeastern University}
        \city{Boston}
        \state{MA}
        \country{USA}}
\email{w.gatterbauer@northeastern.edu}

\author{Prashant Pandey}
\orcid{0000-0001-5576-0320}
\affiliation{%
    \orcidicon{0000-0001-5576-0320}
	\institution{Northeastern University}
        \city{Boston}
        \state{MA}
        \country{USA}}
\email{p.pandey@northeastern.edu}

%% file: abstract.tex
\begin{abstract}
Filters are ubiquitous in computer science, enabling space-efficient approximate membership testing.
Since Bloom filters were introduced in 1970, 
decades of work have improved their space efficiency and performance.
Recently, three new paradigms have emerged that offer orders-of-magnitude improvements in false positive rates (FPRs) 
by leveraging additional information beyond the input set:
(1) \emph{learned filters} train a model to separate members from non-members,
(2) \emph{stacked filters} use negative workload samples to build cascading layers,
and 
(3) \emph{adaptive filters} update internal representation in response to false positive feedback.
Yet each paradigm targets specific use cases, introduces complex configuration tuning,
and has been evaluated only in isolation. This results in unclear trade-offs and a significant gap in understanding 
how these approaches compare and when each is most appropriate.

This paper presents the first comprehensive comparative evaluation of learned, stacked, and adaptive filters
across real-world datasets and diverse query workloads.
Our results reveal critical trade-offs:
(1) \emph{Learned filters} achieve up to $10^2\times$ lower FPRs when query distributions match training data,
but exhibit high variance and lack robustness under skewed or dynamic workloads.
Critically, learned filters' inference overhead 
leads to up to $10^4\times$ slower query latencies than stacked and adaptive filters.
(2) \emph{Stacked filters} reliably achieve up to $10^3\times$ lower FPRs 
on skewed workloads
but require prior knowledge of the workload.
(3) \emph{Adaptive filters} are robust across all settings, 
achieving up to $10^3\times$ lower FPRs under adversarial queries
without workload assumptions.
Based on our analysis, \emph{learned filters} suit stable workloads where input features enable effective model training and space constraints are paramount,
\emph{stacked filters} excel when query distributions are known in advance and relatively static, and
\emph{adaptive filters} are the most general option, 
providing robust guarantees backed by theoretical bounds even for dynamic and adversarial environments.

\end{abstract}

%% file: introduction.tex
\section{Introduction}
\label{SEC:INTRODUCTION}

A \emph{filter} (such as Bloom~\cite{Bloom1970}, quotient~\cite{Bender2012,
Pandey2017}, cuckoo~\cite{Fan2014}, XOR~\cite{Graf2020}, or
ribbon~\cite{Dietzfelbinger2026}) provides an approximate representation of a
set and saves space by trading off accuracy through one-sided errors (no false
negatives) in membership queries. These errors are bounded by a configurable
false positive rate (FPR). 

Filters are ubiquitous across computer science and have long served as
fundamental building blocks in applications where memory constraints are
important and one-sided errors are acceptable~\cite{Dillinger2009, Pandey2018,
  Pandey2017a, Pandey2017b, Broder2004, diaconu2013, Cao2020, Conway2023,
Pandey2020, Nisa2021}. 

Since Bloom's seminal work in 1970~\cite{Bloom1970}, there have been decades of
progress 
in filter design. Dynamic filters such as quotient filters~\cite{Bender2012,
Pandey2017} and cuckoo filters~\cite{Fan2014} improved upon Bloom filters by
supporting deletions and achieving better cache performance. More recently,
static filters such as XOR filters~\cite{Graf2020} and ribbon
filters~\cite{Dietzfelbinger2026} have pushed space efficiency closer to the
information-theoretic optimum.\footnote{The lower bound on space usage for a
dynamic filter is $\opt + \Omega(n)$ bits~\cite{Pandey2017}.} With over a
thousand papers on filter data structures published in the last two decades,\footnote{We searched for Bloom filters on DBLP and got 1,130 results.}  these
advancements have collectively improved both space utilization and query
performance, making filters increasingly efficient for their traditional use
cases.

\begin{figure}[t]
\includegraphics[scale=0.28]{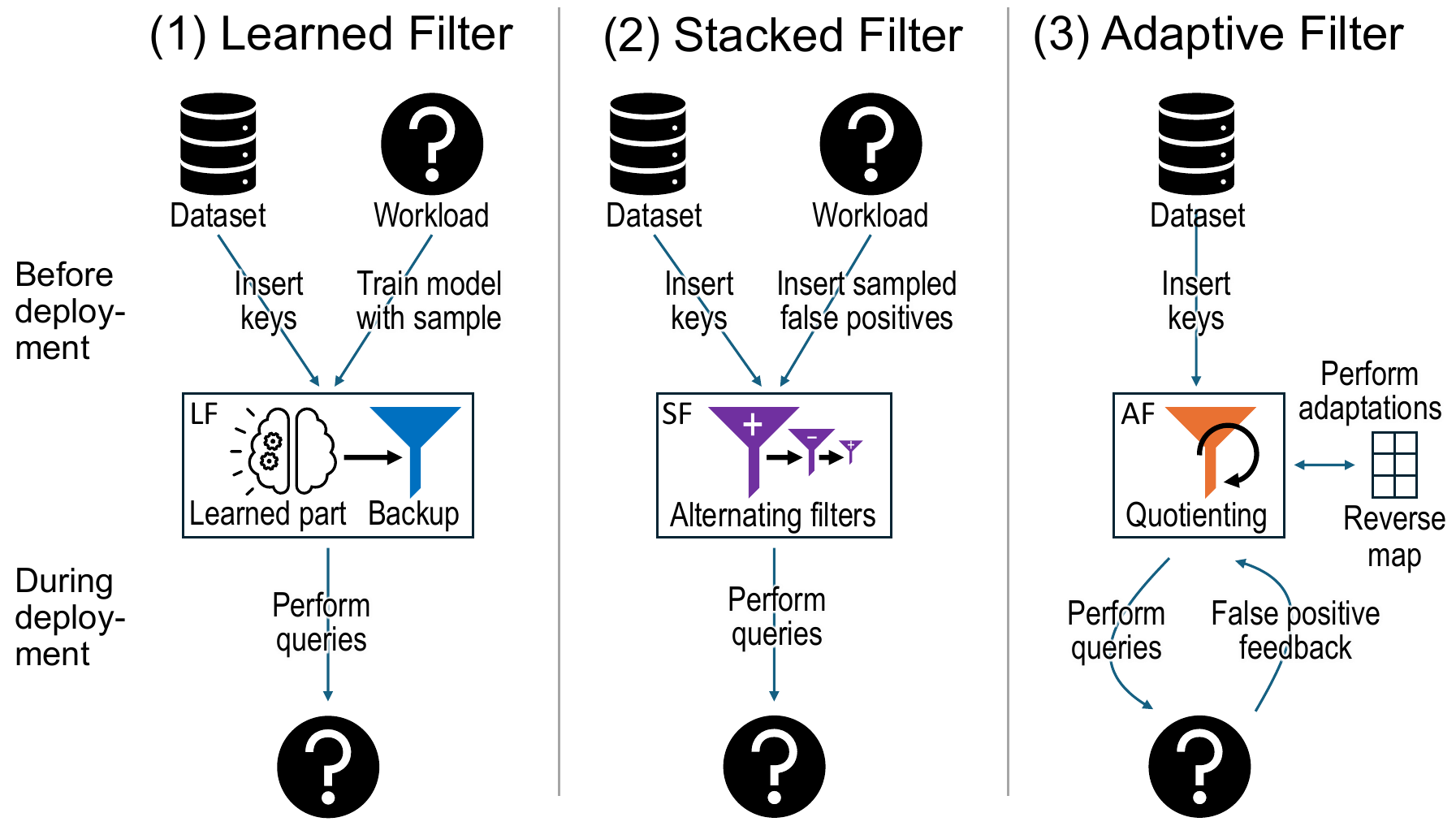}
\caption{
Before deployment, learned filters train on dataset features and stacked
filters train on sampled negative queries. During deployment, adaptive filters
update on false positive feedback.
}
    \label{fig:example-interaction}
    \vspace{-1.5em}
\end{figure}

\para{Modern paradigms in filter design}
Recently, three new paradigms in filter design have emerged that offer
orders-of-magnitude improvements in false positive rates by leveraging
additional information beyond the input set alone. 
\Cref{fig:example-interaction} summarizes these approaches, each optimized for different assumptions about available data, 
workload knowledge, and query feedback.

\begin{enumerate}
\item \textbf{Learned filters}~\cite{Kraska2018, Mitzenmacher2018} exploit
  correlations between element features and dataset distributions by training a
  model to distinguish keys that are not part of the set from those that are.
  Queries first consult the trained model and only proceed to a backup filter
  if the model predicts a negative result. 
  State-of-the-art implementations such as Fast
  Partitioned Learned Bloom Filters (\textsc{PLBF})~\cite{Vaidya2021, Sato2023},
  \textsc{Ada-BF}~\cite{Dai2020}, and Sandwiched Learned Bloom
  Filters~\cite{Mitzenmacher2018} can achieve up to $10\times$ lower false positive
  rates compared to Bloom filters at equivalent space. 

\item \textbf{Stacked filters}~\cite{Deeds2020} leverage workload knowledge from
 query samples to build cascading layers of filters that suppress false positives
  on frequently queried negatives. By storing these frequent positives across
  layers, they often outperform learned filters on skewed workloads.

\item \textbf{Adaptive filters}~\cite{Bender2018, Wen2025, Mitzenmacher2020}
  adjust their internal structure in response to feedback from false positive
  results. By maintaining a disk-resident \emph{reverse map} that recalls the
  original key causing a false positive, adaptive filters can extend their
  structure to avoid repeating the same mistake during future queries. 
  This provides a much stronger guarantee than traditional filters: bounded
  false positive rates for \emph{any} arbitrary sequence of queries, rather
  than only for queries drawn uniformly at random. The state-of-the-art
  adaptive filter, \textsc{AdaptiveQF}~\cite{Wen2025}, achieves up to
  $100\times$ lower false positive rates compared to traditional filters.
\end{enumerate}

\para{Why it is challenging to navigate the design space}
These paradigms introduce several new design knobs (model choice, training
strategy, layer layouts, and feedback policies) that require careful tuning to
achieve desired performance. Furthermore, all three paradigms have been defined
and evaluated primarily within their respective settings, without a unified
evaluation framework to draw meaningful conclusions about how they compare to
one another.

This creates a significant gap in our understanding of modern filter paradigms.
Systems designers face a daunting landscape: several tuning knobs, diverse
application requirements, and disparate performance characteristics. 
At the same time, there is a lack of clear guidance on how to navigate these
choices. Existing learned filter evaluations focus only on stable, predictable
workloads and often compare primarily against Bloom filters, despite stronger
modern baselines (e.g., cuckoo/quotient/xor filters). Critical metrics such as
the computational overhead of model inference and the impact of model sizing
have been largely overlooked, potentially overstating the practical benefits of
learned approaches.

\para{Our contribution}
In this paper, we aim to fill this information gap through a comprehensive
evaluation and comparison of learned, stacked, and adaptive
filters.\footnote{We study approximate membership/equality filters that answer
  `is key $x$ in set $S$?' with no false negatives and tunable FPR. We do not
study range filters, probabilistic predicates beyond membership, or learned
index structures; our focus is the design space of learned, stacked
(workload-trained), and adaptive (feedback-driven) membership filters.} Our
contributions include:

\begin{itemize}
  \item \textbf{Unified evaluation:} We provide the first comprehensive
    comparison of learned, stacked, and adaptive filters under a unified
    evaluation framework, featuring direct performance comparisons across
    diverse workloads and datasets.

  \begin{itemize}
    \item \textbf{Robustness analysis:} We evaluate filter performance under
      adversarial attacks, dynamic workload distributions, and repeated query
      patterns: conditions that stress-test the assumptions underlying each
      filter paradigm.

    \item \textbf{Computational overhead analysis:} We measure and report the
      full computational costs of each filter type, including model training and
      inference overhead for learned filters, constructing cascading layers for
      stacked filters, and adapting and reverse map maintenance for adaptive
      filters.

    \item \textbf{Configuration sensitivity study:} We analyze how key learned filter configuration
    parameters, such as the choice of model, model size allocation, and training
    set size, affect their overall performance.
  \end{itemize}

  \item \textbf{Use case recommendations:} We provide evidence-based
  guidance for selecting appropriate filter types based on application
  requirements, including query distribution stability,
  computational constraints, and robustness needs.
\end{itemize}

\para{Key Findings}
Learned filters demonstrate improved accuracy compared to adaptive and stacked
filters under specific conditions. When query distributions closely match
training distributions, learned filters achieve considerably better FPRs
compared to the other filters (close to $10^2 \times$ better
(\cref{fig:one-pass-results})). However, learned filters can suffer from high
variance in observed false positive rates and lack robustness guarantees
against adversarial attacks (\cref{fig:adversarial-results}) and dynamic
workloads (\cref{fig:dynamic-rebuild}), even when model rebuilding is allowed. 
In particular, for skewed workloads, learned
filter false positive rates can even fluctuate from $1\%$ to over $40\%$ when
\textit{increasing} the space budget (\cref{fig:zipfian-results}).

Critically, our analysis demonstrates that the computational cost of querying
learned models frequently exceeds that of traditional filter operations (up to
$10^4 \times$ slower (\cref{fig:overall-query})), challenging the practical
value proposition of learned filters in many applications. This finding is
particularly significant given that existing evaluations have largely
overlooked model querying costs, potentially overstating the benefits of
learned approaches.

Stacked filters, by contrast, often achieve lower false positive rates
across diverse datasets and query patterns for equivalent space allocations
because of their use of workload knowledge. For example, on queries following a
Zipfian distribution, they can reach false positive rate improvements of up to
$10^5 \times$ (\cref{fig:zipfian-results}) compared to some variants of learned
filters.

Adaptive filters provide robust theoretical guarantees and demonstrate low
false positive rates across different datasets and query patterns. On workloads
where the query distribution changes adversarially, the adaptive filter's false
positive rates can be up to $10^3 \times$ (\cref{fig:adversarial-results})
lower than other filters.
Adaptive filters also offer the fastest query and construction performance, beating
stacked filters (the next best filter) by up to $3\times$.
Adaptive filters demonstrate dependable
performance regardless of dataset characteristics, consistently decreasing the
false positive rate as more space is allocated (\cref{fig:one-pass-results},
\cref{fig:uniform-results}, \cref{fig:zipfian-results}).

We also demonstrate new findings regarding learned filter configurations. We
show that the proportion of negative keys in the training set has little effect
on the final learned filter performance, with median false positive rates
having ranges close to $0$ across different proportions
(\cref{fig:model-degrad}). We also find that assigning too much relative space
to the trained model within a learned filter's space budget has adverse effects
on performance, with the filters exhibiting false positive rates well over
$10\%$ when the majority of the space is dedicated to the model
(\cref{fig:changing-model}).

Through validation of theoretical results and comprehensive experimental
analysis, this work establishes specific application requirements to guide a
choice between learned, stacked and adaptive filters. With their high space
efficiency on predictable workloads, learned filters are suitable for stable
workloads where the features of input keys can be meaningfully used to improve
model decisions and inference cost overheads are not critical. Meanwhile, the
robustness and higher-speed operations of adaptive and stacked filters make
them useful for general settings, with benefits in particular for dynamic,
security-sensitive, or high-workload applications. Adaptive filters
specifically are not constrained by knowledge about the query workload and are
the most generalizable filter.

%% file: learnedfilters.tex
\section{Learned filters}
\label{SEC:LEARNEDFILTERS}

A \emph{learned filter} combines a machine learning model with a traditional filter (usually a Bloom filter \cite{Dai2020, Sato2023}) to improve the space-accuracy trade-off for approximate membership testing.
The learned filter operates in two stages during deployment:
\begin{enumerate}
  \item \textbf{Model prediction.} When a query arrives, the filter takes extracted features 
  from the key and passes them to a trained machine learning model. The model outputs a confidence score $s \in [0,1]$ 
  indicating the likelihood that the key belongs to the positive set.

  \item  \textbf{Threshold decision.} If the score exceeds a predetermined 
  threshold $t$ (meaning $s \geq t$), 
  the filter immediately returns "positive". If the score is below the threshold $(s < t)$, the query is passed to a backup traditional filter for the final decision.
  
\end{enumerate}

\para{Training process}
The model is trained on a dataset containing both positive examples (keys that should return "yes") and negative examples (keys that should return "no"). 
For a given key, features suitable as ML model input can either be extracted from the key or given as separate input.
The training process learns correlations between features of keys and their membership status. 
For example, in malicious URL detection, features might include URL length, domain characteristics, and character patterns \cite{features:url}. In malware detection, features could include  file size, system versions, and other important metadata \cite{2018arXiv180404637A}.

\para{Key insight}
The learned model acts as a "smart pre-filter" that identifies likely positives based on learned patterns, allowing the backup filter to be smaller since it only handles the remaining uncertain cases. When the model correctly identifies positives, it saves space because those keys do not need to be stored in the backup filter. The backup filter also ensures the overall structure maintains the guarantee of a $0\%$ false negative rate even if the model makes a mistake.

\para{Critical dependency}
The improved effectiveness of learned filters strongly depends on the quality of the underlying machine learning model and the assumption that query patterns will match training patterns. When these assumptions break down
(as often happens in practice)
the filter's performance can degrade significantly, sometimes performing worse than traditional approaches.

Note that for a learned filter, false positives from the ML model due to high scores become false positives in the overall output. 
Thus, the FPR of the learned filter is floored by the FPR of its ML model. Concretely, if the model has a FPR $f_m$ 
and its backup filter structure has a FPR $f_b$, then the overall FPR of a learned filter is $f_m + (1-f_m)(f_b)$~\cite{Mitzenmacher2018}. 
Due to their dependence on model quality, learned filters tend to be used in cases where there is a direct correlation between the features of an element and the likelihood of its presence in a set.

Mitzenmacher~\cite{Mitzenmacher2018} also introduces the \emph{sandwiched learned Bloom filter}. 
In a sandwiched learned Bloom filter, an additional filter is employed before the trained model 
to process queries and immediately return true negatives.
This additional step reduces the number of potential false positives that the rest of the learned filter processes.

\subsection{Theoretical guarantees}
\textbf{Behavior of learned models.} 
Mitzenmacher~\cite{Mitzenmacher2018} used a Chernoff bound to prove a theorem
suggesting that the empirical FPR of a learned filter on a test set will be close to the FPR on future queries 
\textit{assuming that the test set has the same distribution as future queries}. 
We will experimentally demonstrate that when this assumption does not hold, 
learned filters cannot strongly guarantee their performance, resulting 
in substantial deviations in empirical FPRs.

\textbf{Size of learned models.} 
Analysis from \cite{Mitzenmacher2018} also describes how the size of the learned model relates to its expected effectiveness compared to standard Bloom filters.
 \begin{theorem}{\cite{Mitzenmacher2018}}
    Consider a learned Bloom filter
    with $n$ keys, 
    $b$ bits per key used by the backup filter, and 
    slot fill rate $\alpha$. 
    For its learned function using $\zeta$ bits, let $f_m$ be the false positive probability, and $f_n$ the false negative probability. 
    The learned filter is expected to outperform a Bloom filter using the same space budget when
    $$\zeta / n \le \log_\alpha(f_m + (1-f_m)\alpha^{b/f_n})-b$$
    \label{thm:space}
    \vspace{-1.5em}
\end{theorem}

A bound can be applied to the number of bits per key the learned function should be allotted to outperform standard filters. 
Intuitively, as the amount of space dedicated to the learned function increases, it has a reduced benefit from the backup filters, eventually worsening its performance. 
We will experimentally demonstrate how changing the proportion of space dedicated to the learned function influences the performance of the overall learned filter.

\begin{figure}[t]
\includegraphics[width=0.9\linewidth]{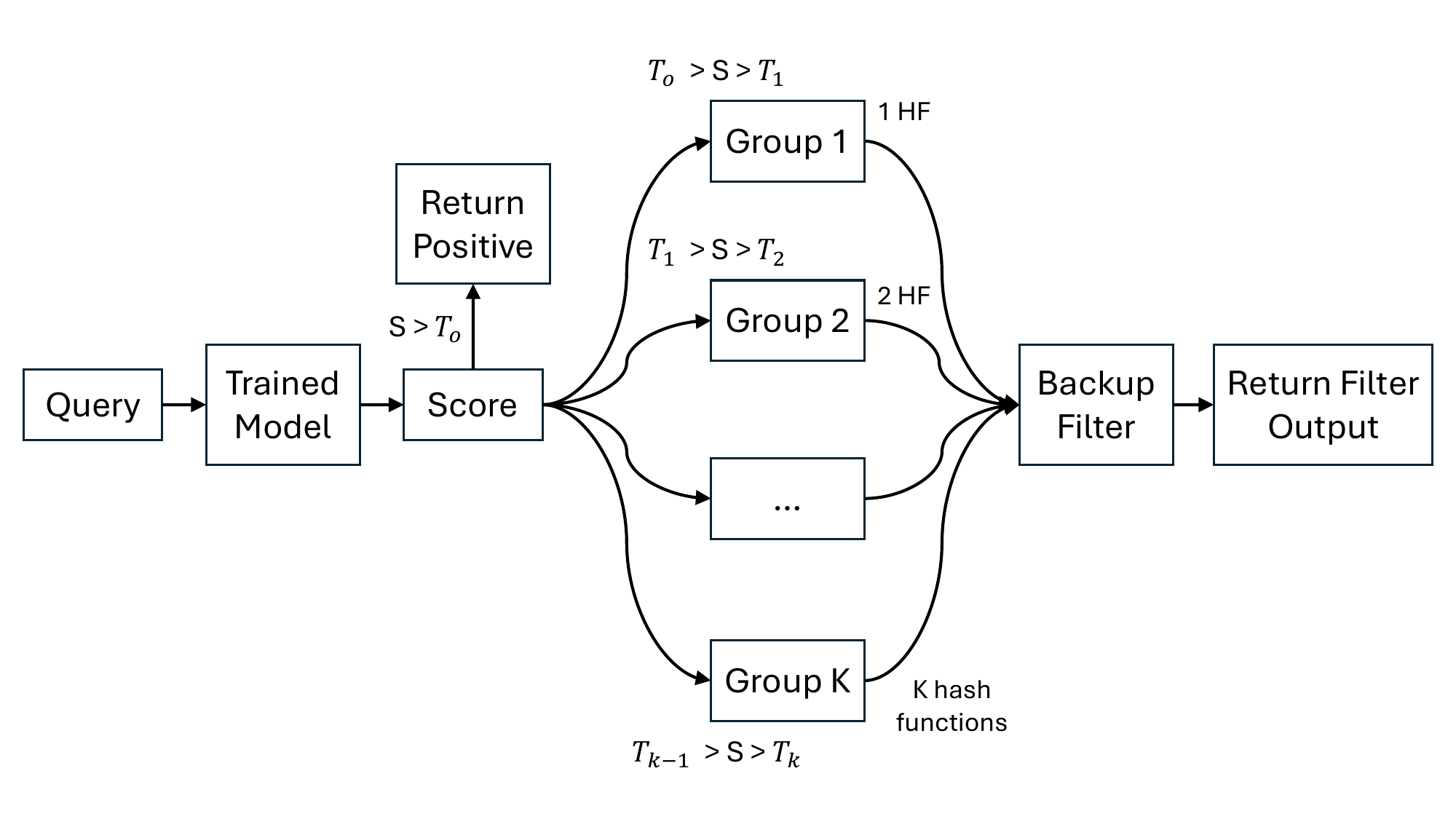}
\caption{\textsc{Ada-BF} structure. Keys are assigned a number of hash functions depending on the model's predicted likelihood of them being a positive element. The bits corresponding to all their hashes are used for insertions and queries.
}
\label{fig:adabf}
\end{figure}

\subsection{Ada-BF}
\textsc{Ada-BF} expands on the basic learned Bloom filter by adapting insertions and queries into the backup filter according to the internal model's predictions \cite{Dai2020}. The structure is described in \cref{fig:adabf}.

First, it splits queries into $k$ groups. Group assignments depend on the model's score for the input key, which describes the likelihood that the key is a positive element. The score thresholds to divide the groups are optimized according to a constant $c$ such that each group is expected to contain $c$ times more non-keys than the next group. Within each group, keys are assigned a certain number of hash functions. Groups with more keys are assigned increasingly larger numbers of hash functions, up to $k$.

During an insertion, a key is assigned a group according to its model score, then all the bits in the backup filter corresponding to its hashes are set. A query checks those same bits if the model predicts that the query is a negative key. Intuitively, to avoid false positives, groups with more keys should check more bits in the backup filter to reduce the chances of a hash collision.

\begin{figure}[t]
\includegraphics[width=0.9\linewidth]{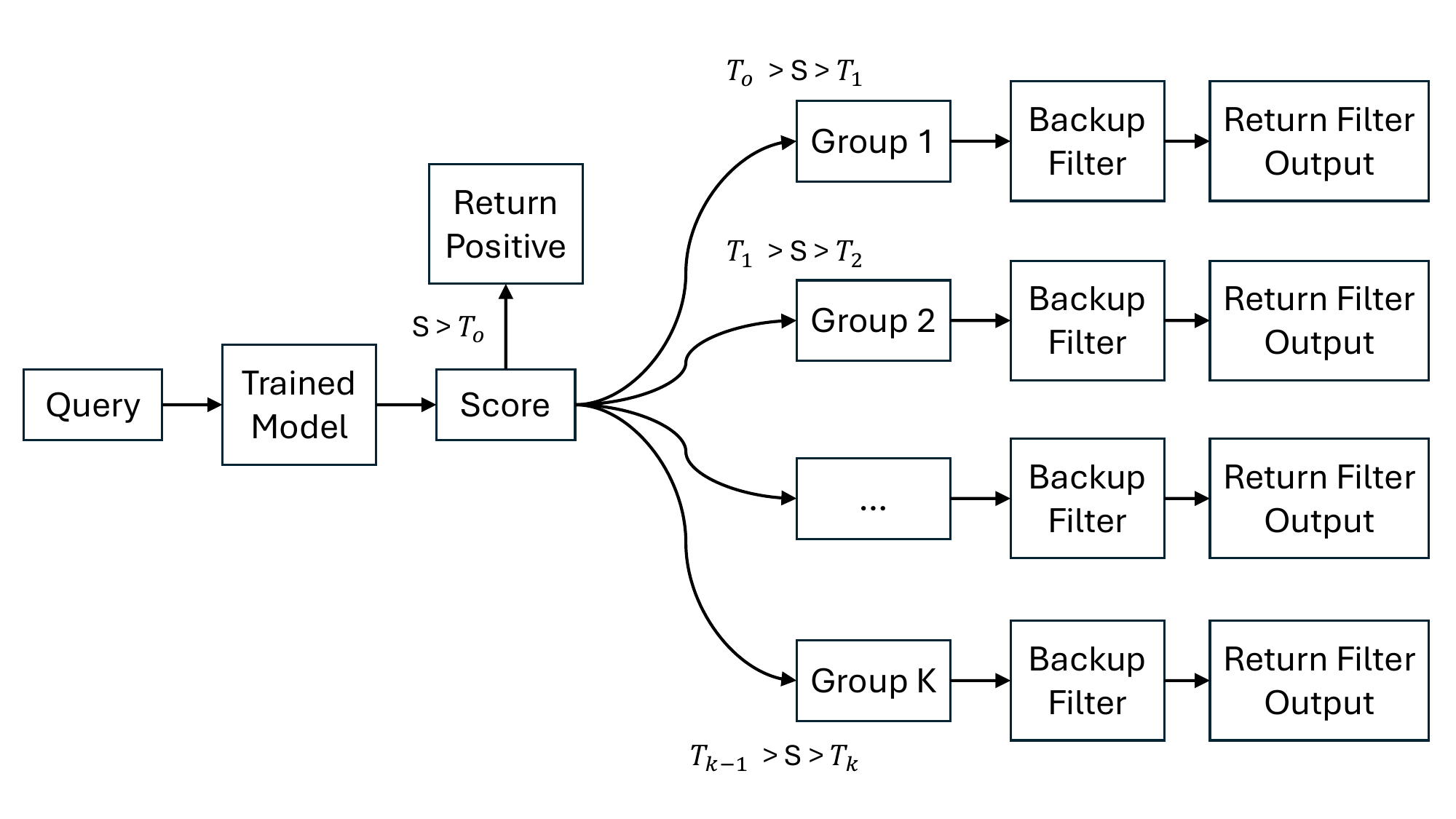}
\caption{Partitioned learned Bloom filter (\textsc{PLBF}) Structure. Keys are assigned to different backup filters depending on the score assigned to them by a model.
}
\label{fig:plbf}
\end{figure}

\subsection{PLBF}
The Partitioned Learned Bloom Filter (\textsc{PLBF}) expands on the basic learned Bloom filter by assigning keys to different backup filters depending on their model scores. The structure is described in \cref{fig:plbf}.

Similarly to \textsc{Ada-BF}, the \textsc{PLBF} divides keys into $k$ groups based on their assigned scores from an internal ML model. Each group has its own backup filter, which is optimized for the space budget and targeted overall FPR. In our experiments, we use the \textsc{Fast PLBF} implementation~\cite{Sato2023}, which uses a dynamic programming approach to efficiently construct the filter by splitting the score space into $N$ segments which are assigned to $k$ groups.

When a key is inserted, it is assigned a group according to its model score, then inserted into that group's backup filter. On a query, if the model predicts the key may be a negative, the filter belonging to the key's group is checked. The intuition is that keys that are likely to be negative should use larger backup filters to reduce their false positives; meanwhile, keys with higher scores use smaller backup filters, since false positives on keys which were likely true positives have less impact on the overall FPR.

%% file: stackedfilters.tex
\section{Stacked filters}
\label{SEC:STACKEDFILTERS}

Stacked filters~\cite{Deeds2020} exploit knowledge of the query workload to
reduce false positive rates. By storing frequently queried negative keys across
cascading filter layers, the structure adapts to the distribution of the query
set. 

A stacked filter consists of alternating layers that store either positive or
negative keys. Construction proceeds as follows: the first layer stores all
positive keys (keys that are part of the set). From a sample of negative
queries, any false positives of the first layer are stored in the second layer.
Any positive values that collide with the second layer are then stored in the
third layer, and so on. This cascading process continues until the space budget
is exhausted or the target false positive rate is achieved.

\begin{figure}[t]
    \includegraphics[width=1.0\linewidth]{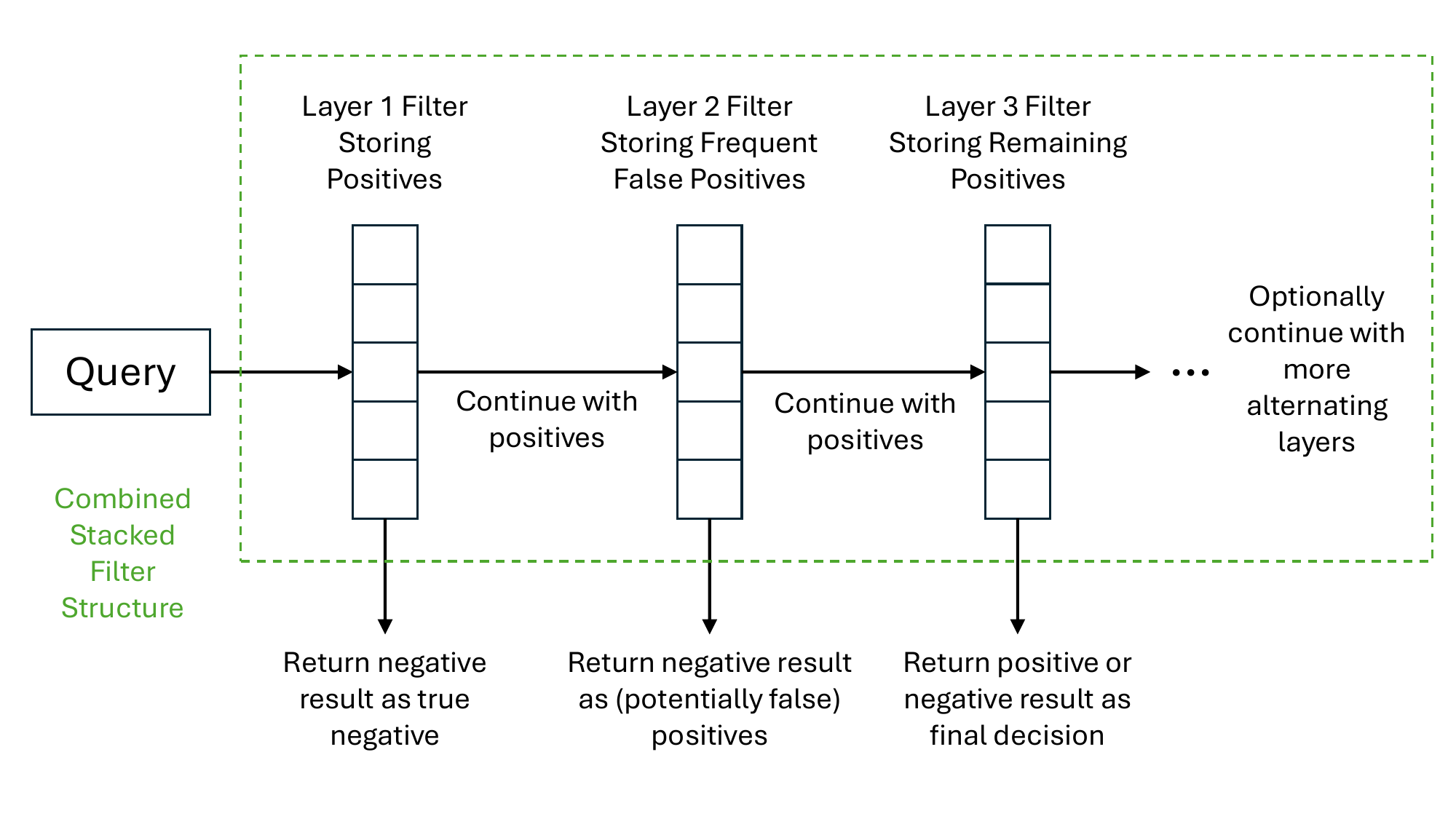}
    \caption{Stacked filter structure. Queries undergo membership queries across layers of filters alternately storing positive and negative queries.
    }
    \label{fig:stackedfilter}
    \vspace{-1.5em}
\end{figure}

Queries traverse the layers sequentially. At each positive layer, a negative
response confirms the element is not in the set. 
At each negative layer, a
negative response indicates the element is likely in the set. 
The query
continues through successive layers until a definitive answer is reached or the
final layer makes the decision. \Cref{fig:stackedfilter} illustrates this
process on a three-layer stacked filter.

The key insight is that frequent negative queries that would cause false
positives in a standard filter are explicitly stored in the stacked filter's
negative layers, eliminating these errors. For infrequent negatives, the
cascading structure provides protection: an element must pass membership checks
across multiple alternating layers, making false positives exponentially
unlikely.

Stacked filter construction requires a sample of the query workload to estimate the
distribution of negative queries. Given this sample and a space budget, the
stacked filter solves an optimization problem to determine the false positive
rate for each layer that minimizes the overall false positive rate. The
structure is flexible, each layer can use any underlying filter type,
including Bloom, quotient, or cuckoo filters.

The expected false positive rate combines two components: the probability that
frequent negatives escape the negative layers, and the probability that
infrequent negatives pass through all layers. Space usage is the sum of filter
sizes across all layers, dominated by the first layer which must store all
positive values. 

The stacked filter also supports incremental construction through a variant
called the adaptive stacked filter~\cite{Deeds2020}, which adjusts to the query
workload online, provided the query distribution remains relatively stable.
This variant initializes with a single layer containing all positive elements.
As queries arrive, detected false positives are inserted into a second
(negative) layer. Once this layer reaches a target load factor, the filter
rescans the positive values to construct a third layer, and subsequent layers
are built similarly as needed.

Though adaptive, the adaptive stacked filter lacks theoretical guarantees and
empirical performance compared to the \textsc{AdaptiveQF}. First, adapting to a
false positive requires substantially more space: the stacked filter must store
the entire false positive key in a new layer, whereas the \textsc{AdaptiveQF}
only extends the colliding fingerprint. Second, the \textsc{AdaptiveQF}
provides stronger theoretical guarantees on false positive rates that the
adaptive stacked filter lacks. Third, query performance differs substantially:
stacked filters probe multiple layers, incurring multiple cache misses, while
the \textsc{AdaptiveQF} guarantees single cache-line access for both positive
and negative queries, yielding significantly better empirical performance.

\para{Similarity to cascade filters}
Stacked filters~\cite{Deeds2020} share conceptual similarities with cascading
filters~\cite{Salikhov2014}, which were introduced in 2014 for representing de
Bruijn graphs in genomics. In this application, the filter represents the set
of $k$-mers (length-$k$ sequences) forming nodes in the graph, where edges
connect nodes sharing a $(k-1)$-length overlap. Rather than storing edges
explicitly, the filter is queried for all possible extensions of a
node, typically bounded by the alphabet size (four nucleotides in genomics).
Since graph traversal defines the query set in advance, cascading filters
construct successive layers of exponentially decreasing size, each storing
false positives from the previous layer, ultimately achieving an exact
representation.

%% file: adaptivefilters.tex
\begin{figure}[t]
    \centering
    \includegraphics[width=\linewidth]{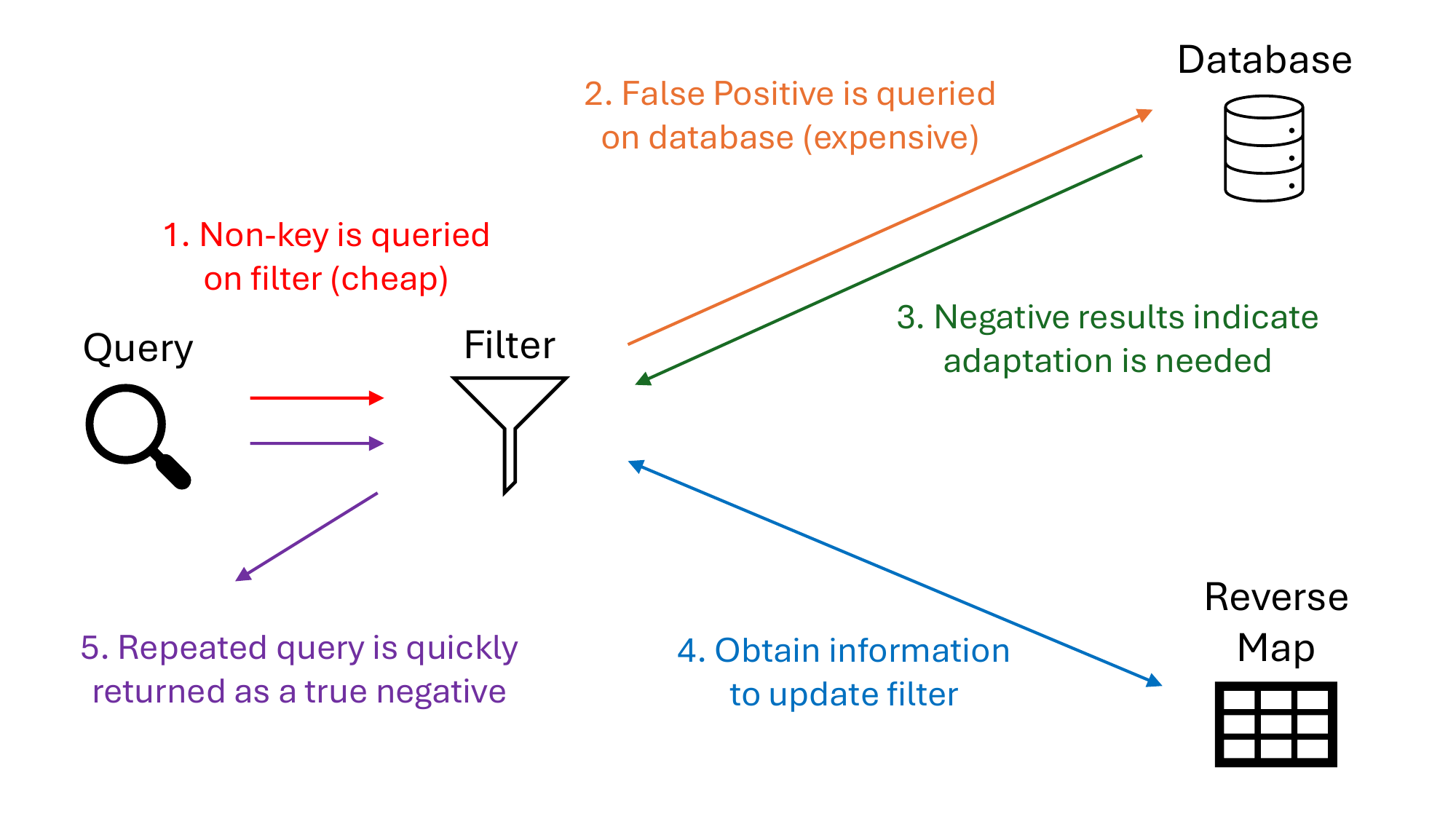}
    \caption{
    Adaptive filters update their representation upon detecting a false
    positive to avoid committing the same false positive again. To adapt they
    employ an on-disk reverse map to access the original key corresponding to
    the colliding fingerprint. This reverse map is only accessed upon false positives.
    }
    \label{fig:filter-example}
    \vspace{-1.5em}
\end{figure}

\section{Adaptive filters}
\label{SEC:ADAPTIVEFILTERS}

An adaptive filter returns \textsc{True} with probability at most $\epsilon$ for every negative query, regardless of answers to previous queries. 
These were defined by Bender et al.~\cite{Bender2018, BenderDaFa21}, who outlined the filter's bounds on the number of false positives (even with a skewed or adversarial query distribution) then presented the broom filter which meets their definition. 
For a database using an adaptive filter, any sequence of $|Q_N|$ negative queries results in $O(\epsilon |Q_N|)$ false positives with high probability. 
This gives a strong guarantee on the number of (expensive) negative accesses that the database will need to make to disk.

Adaptive filters require two features to support adaptivity. 
Firstly, they need \emph{feedback} about their false positives.
For example, if an adaptive filter is used by an application to avoid database lookups for nonexistent items, 
then the application can inform the filter that a query for an item $x$ was a false positive if the 
subsequent database query returned that $x$ is not present. 
Secondly, adaptive filters also need a \emph{reverse map} to store information to support adaptation. 
Bender \etal~\cite{Bender2018} showed that the reverse map is necessary and must be large enough to store the entire set: 
the total size of an adaptive filter on a set $S$ essentially must be large enough to store $S$~\cite{Bender2018}.  
An example is provided in~\Cref{fig:filter-example}. 

The trick is to break the filter into two parts: a small in-memory component accessed on every query 
and a large reverse map accessed only during adaptations (and hence can reside in slower storage). 
All proposed adaptive filters use this structure. 
In some applications, like when the filter is in front of a database, the database may also serve as the reverse map, 
so that the total storage requirements of the system remain essentially unchanged~\cite{Wen2025}.

Later, Mitzenmacher et al.~\cite{Mitzenmacher2020} proposed adaptive cuckoo filters and Lee et al.~\cite{Lee2021} proposed telescoping filters as implementations of adaptive filters to address the skewed query distribution problem. However, these initial implementations suffer from suboptimal performance due to the inefficient design of the reverse map.

Recently, Wen et al.~\cite{Wen2025} proposed \textsc{AdaptiveQF}, a practical implementation of the adaptive filter based on the quotient filter. \textsc{AdaptiveQF} is \emph{monotonically adaptive}, i.e., it never forgets a false positive once adapted. The filter can be resized in case it runs out of space during adaptations and offers strong performance guarantees. 
We employ the adaptive quotient filter (\textsc{AdaptiveQF}) as the representative adaptive filter as it is the state of the art in terms of performance and theoretical guarantees.

\subsection{Adaptive quotient filter}
The \textsc{AdaptiveQF} makes use of \emph{quotienting}. 
For an element $x$, after obtaining its hash $h(x)$, the higher-order $q$ bits are used to obtain a quotient $h_0(x)$ and the lower-order $r$ bits form its remainder $h_1(x)$.
Together, the quotient and remainder make up a $q+r=p$-bit fingerprint.
The filter is initialized by allotting $2^q$ slots for remainders, each with an initial size of $r$ bits. 
During an insertion of a key $x$, the filter first obtains $h_0(x)$ and $h_1(x)$ from its hash. 
$h_0(x)$ is then used to assign the slot to store $h_1(x)$. 
To query a key, the filter checks if any of the stored elements in the slot corresponding to the queried key's quotient matches its remainder. 

A false positive occurs if the $p$-bit fingerprint (quotient and remainder) of an existing key matches one that is not in the set. 
After a false positive query, the stored remainder is extended by the next $r$ bits in the hash of the original key. 
The \emph{reverse map} stores the mappings of fingerprints to their original inserted key. 
This mapping allows the filter to find the original key corresponding to the collided fingerprint and determine which bits should be used to extend the fingerprint by recreating the complete hash. 
By `over adapting' and adding $r$ more bits instead of the minimum required (in expectation 2), the adaptive filter ensures that for some specified FPR $\epsilon$, it never makes errors above that threshold~\cite{Wen2025}. 

The original paper demonstrates that with its performance guarantees, 
the \textsc{AdaptiveQF} implementation performs empirically better than 
other adaptive filters and remains competitive with non-adaptive filters 
while still maintaining its FPR \cite{Wen2025}. 
The main draw of this filter is its monotonic adaptivity guarantees, 
allowing its FPR to be dependable for arbitrary query distributions.

%% file: experiments.tex
\section{Experimental methodology}
\label{SEC:EXPERIMENTS}

In this section, we extensively evaluate the performance of state-of-the-art
learned, adaptive, and stacked filters. Our evaluation aims to fill the gaps in
our understanding of the performance of these filters 
\iflongversion
as pointed out
in~\Cref{sec:evaluation-gaps}
\fi
.

Furthermore, we perform an investigation into learned filters to better
understand the impact of various configuration parameters.
We assess the impact of considerations like the ratio of the model and backup
filter(s) sizes or the composition of the model's training set on the
structure's overall false positive rate.

Our evaluation represents the first direct comparative analysis between
learned, adaptive, and stacked filtering paradigms, providing evidence-based
guidance for filter selection in practical deployments.

\para{Filters evaluated}
We evaluate 
implementations from each modern paradigm (learned,
stacked, adaptive). 
As representatives, we compare the \textsc{Ada-BF}~\cite{Dai2020},
\textsc{PLBF}~\cite{Vaidya2021}, stacked filters~\cite{Deeds2020},
and the \textsc{AdaptiveQF}~\cite{Wen2025}. We exclude the (sandwiched) learned
Bloom filter because \textsc{Ada-BF}, \textsc{PLBF}, and stacked filters all
demonstrated superior empirical performance under direct
comparisons~\cite{Dai2020, Vaidya2021, Sato2023, Deeds2020}. We use the
\textsc{Fast PLBF} implementation instead of \textsc{PLBF++} because under some
conditions the \textsc{PLBF++} differs in structure from the \textsc{PLBF}
\cite{Sato2023}. For all the filters, we employ the open source implementations
provided by the original authors. The learned filters~\cite{Dai2020, Sato2023}
are implemented in Python due to the library support for model
training~\cite{Pedregosa2011}, the \textsc{AdaptiveQF}~\cite{Wen2025} is
implemented in C language, and the stacked filter~\cite{Deeds2020} is
implemented in C++.

\para{Implementation language differences} While the choice of language for
filter implementation such as Python or C can introduce differences in run
time, we choose to employ the learned filter implementations provided by the
original authors due to two main reasons. First, we show
in~\cref{fig:overall-query} that for learned filters, the cost of model
inference is orders of magnitudes larger than the cost of querying backup
filters, contributing to a performance difference too large to be attributed
solely to language differences. Second, we want to avoid any author bias when
implementing them ourselves in a different language. 
Our goal is to evaluate the
open-source implementations provided by the original authors.

\para{FPR calculation} We also define the reported false positive rate (FPR) in
terms of the content of the query set. After performing a set of queries $Q$ on a
filter, 
if $Q_{FP} \in Q$ is the multiset of false positives and
$Q_N \in Q$ is the multiset of negatives,
then the empirical false positive rate is $FPR = \frac{|Q_{FP}|}{|Q_{FP}|
+ |Q_{N}|}$.

\subsection{Research questions}

Our experimental evaluation aims to address the following questions regarding the relative performance of 
modern filters:

\begin{enumerate}
    \item How do learned, stacked, and adaptive filters compare in FPR and
      space efficiency under identical workload conditions?
    
    \item How do learned, stacked, and adaptive filters perform under
      real-world query distributions (uniform, Zipfian, adversarial) versus the
      stable workloads assumed in current evaluations?
    
    \item What are the true computational costs, such as construction time,
      query throughput (including model inference), and retraining overhead, of
      learned filters compared to stacked and adaptive filters?
    
    \item How do learned filter configuration choices (space allocation, model
      type, training set composition) affect performance, and do empirical
      results validate existing theoretical bounds?
    
    \item How do learned filters perform on general-purpose datasets lacking
      strong feature-membership correlations, beyond the specialized domains
      (malware, URL filtering) used in current evaluations?
    
    \item How do learned, stacked, and adaptive filters degrade under dynamic
      workloads with evolving datasets and shifting query patterns, and what
      maintenance strategies sustain performance over time?

  \end{enumerate}

\subsection{Evaluation framework}

To address these questions, we conduct a comprehensive experimental evaluation across multiple real-world datasets, 
diverse query distribution patterns, various performance metrics, and alternative learned model configurations. 
All code is available \href{https://github.com/saltsystemslab/How-to-Train-Your-Filter}{online}.

\para{Machine specification}
All experiments are run on a server with an Intel(R) Xeon(R) Gold 5218 CPU @ 2.30GHz using Linux kernel 5.14.0 and 754 GiB RAM.

\subsection{Datasets}
We use the real-world Malicious URL, Ember, Shalla, and Caida datasets from the papers that proposed these filters~\cite{Dai2020, Sato2023, Wen2025}. 
The datasets are described in \cref{tab:dataset-summary}.
Shalla~\cite{dataset:shalla} and Caida~\cite{dataset:caida} 
are new for learned filter evaluation, 
while Malicious URL~\cite{kaggle-url} and Ember~\cite{2018arXiv180404637A} are new for adaptive filters.
Stacked filters were previously tested on a variant of the Shalla dataset \cite{Deeds2020}.

For the Malicious URL dataset~\cite{kaggle-url}, we insert URLs labeled as "malicious" into the filter, 
while all other URLs are treated as negative keys. 
For the Ember dataset~\cite{2018arXiv180404637A}, we insert the SHA-$256$ hashes of malware into the filter. 

 For the Shalla dataset~\cite{dataset:shalla}, since the blacklist consists only of malicious URLs, 
 we combine the dataset with the Cisco Umbrella Top 1M Domains~\cite{dataset:cisco}, 
 treating any URLs that are not present in the Shalla blacklist as negative keys.

The CAIDA dataset \cite{dataset:caida} consists of a set of anonymized network traces. 
To convert the dataset into a classification problem, we filtered the dataset for entries that only followed 
the TCP or UDP protocols, then treated traces following the UDP protocol as positive keys.

\begin{table}[h]
\centering
\small  %
\begin{tabularx}{\columnwidth}{|X|X|X|X|X|}
 \hline
 Dataset Name & Overall Keys & Positive Keys \\
 \hline
 URL & \phantom{0,}162,798 & \phantom{0,0}55,681 \\
 Ember & \phantom{0,}800,000 & \phantom{0,}400,000 \\
 Shalla & 3,905,928 & 2,926,705  \\
 Caida & 8,493,974 & 1,196,194 \\
 \hline
\end{tabularx}

\caption{Summary of datasets.}
\label{tab:dataset-summary}
\vspace{-3.5em}
\end{table}

\subsection{Filter construction}
\label{sec:filter-const}

For all datasets, we start by constructing an
\textsc{AdaptiveQF}~\cite{Wen2025}. Recall that the \textsc{AdaptiveQF} has two
parameters $q$ and $r$ that control the filter's space usage and FPR
(\Cref{SEC:ADAPTIVEFILTERS}). 
If $n$ is the number of unique positive key elements to insert from
the dataset, then we allocate $2^q$ slots for all insertions by setting
$q=\lceil\log_2(n)\rceil$.
We next vary $r \in \{5, 6, 7, 8, 9, 10, 11\}$ and keep track of the seven
possible sizes of the \textsc{AdaptiveQF} in each dataset.

For each dataset, we construct an \textsc{Ada-BF}~\cite{Dai2020} and Fast
\textsc{PLBF}~\cite{Sato2023} by taking the corresponding size of the
\textsc{AdaptiveQF} and subtracting the size of the trained model to obtain the
space budget of the backup filters. For \textsc{Ada-BF}, we set
$k_{min}=8$,$k_{max}=11$, $c_{min}=2.1$, and $c_{max}=2.6$, while for Fast
\textsc{PLBF} we use $N=1000$ and $k=5$.

We also construct a stacked filter \cite{Deeds2020} matching
the size of the \textsc{AdaptiveQF}. During construction, the stacked filter is
given the first $25\%$ of queries and their true results to learn the
distribution of frequently occurring negatives.

This setup allows us to directly compare the FPR-space trade-off for each filter. 
Although each subsequent learned filter for a dataset has a smaller proportion of space 
dedicated to its internal model, we can now evaluate the effectiveness of the usage of the filters 
rather than having a confounding variable in the potentially improving accuracy of the model 
as its allowed space grows.

\para{Adaptive filter reverse map}
The adaptive filter requires a reverse map to recover original keys during
adaptation. For an insertion set of size $n$, the reverse map is $O(n)$ in size and is
therefore stored on disk and is accessed only upon false
positives~\cite{Wen2025}. In production deployments, the backing database
itself typically serves as the reverse map: since false positives already
trigger disk accesses to verify membership, adaptation adds negligible overhead
by reusing this lookup~\cite{Wen2025}.

To isolate in-memory filter performance from disk I/O effects, all our experiments use
in-memory implementations, including an in-memory chaining hash table for the
reverse map. This ensures runtime measurements reflect actual filter operation
costs rather than disk latency.

\begin{figure*}[h]
    \centering
    \includegraphics[width=\linewidth]{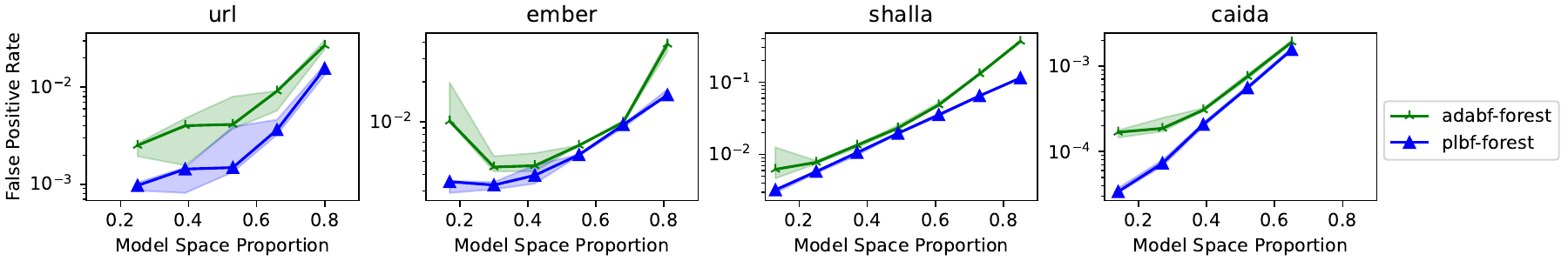}
    \caption{Effect of increasing proportion of (fixed) filter space dedicated to model on FPR (10M uniform queries on each dataset).
    Increasing the space usage of the model in a learned filter corresponds to an increased FPR.}
    \label{fig:changing-model}
\end{figure*}

\begin{figure*}[h]
    \centering
    \includegraphics[width=\textwidth]{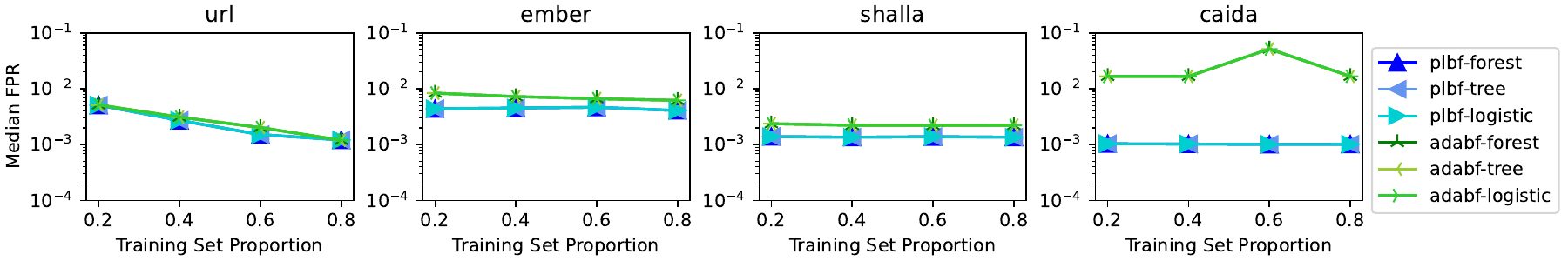}
    \caption{Effect of train set negative key proportion on median FPR (10M uniform queries on each dataset).
    Little correlation is shown between the proportion of negative keys used for training and the learned filter's preformance.}
    \label{fig:model-degrad}
\end{figure*}

\subsection{Training learned filters}
\label{sec:trained learned filters}
Between trials, a new model is trained before constructing the filters, with all learned filters using the same model. 
We start by comparing the size and accuracy of \texttt{sklearn}'s \texttt{RandomForestClassifier}, \texttt{DecisionTreeClassifier}, 
and \texttt{LogisticRegression} models~\cite{Pedregosa2011}.
The logistic regression models are expected to be the smallest, followed by the decision trees, with random forests serving
as the largest models.

The parameters were initially chosen by performing a grid search on a set of
candidate parameter values which were expected to fit under the size of the the
smallest \textsc{AdaptiveQF} (guaranteeing space for backup filters), scoring
based on the balanced accuracy across positive and negative values. Some grid
search results still exceeded the size of the smallest \textsc{AdaptiveQF}, so
those parameters were manually adjusted towards smaller values.
Each dataset has different values for the model parameters, as described in \cref{tab:parameter-summary}.

\begin{table}[h]
\centering
\small  %
\begin{tabularx}{\columnwidth}{|X|X|X|X|X|}
 \hline
 Dataset Name & Estimators (Forest) & Leaf Nodes (Forest) & Leaf Nodes (Tree) & C (Logistic)\\
 \hline
 URL & 30 & \phantom{0,0}10 & \phantom{0,}320 & \phantom{0000}0.1 \\
 Ember & 10 & \phantom{0,}320 & 1,280 & 0.00001\\
 Shalla & 20 & 1,280 & 1,280 & \phantom{0.000}10 \\
 Caida & 10 & 1,280 & 1,280 & 0.00001 \\
 \hline
\end{tabularx}
\caption{Summary of model parameters for datasets.}
\label{tab:parameter-summary}
\vspace{-2.5em}
\end{table} 

From a dataset, we randomly draw $30\%$ of the negative keys and all of the positive keys as the training set, 
then use all of the negative keys and positive keys as the testing set. 
Example model sizes and accuracies from one trial of model training using the corresponding parameters from~\cref{tab:parameter-summary} are
included in~\cref{tab:model-summary}, which indicate that lower model sizes correspond to lower accuracy.

{
\setlength{\tabcolsep}{3pt}
\begin{table}[h]
\centering
\small  %
\begin{tabular}{|l|cc|cc|cc|}
 \hline
 Dataset & 
 \multicolumn{2}{c|}{Random Forest} & 
 \multicolumn{2}{c|}{Decision Tree} & 
 \multicolumn{2}{c|}{Logistic Regression} \\
 \cline{2-7}
 & Size (KB) & Accuracy
 & Size (KB) & Accuracy
 & Size (KB) & Accuracy \\
 \hline
 URL & \phantom{0,0}67 & 0.94 & \phantom{0}52 & 0.96 & \phantom{0}1 & 0.90  \\
 Ember & \phantom{0,}515 & 0.96 & 206 & 0.97 & 10 & 0.53 \\
 Shalla & 4,102 & 0.7 & 206 & 0.71 & \phantom{0}1 & 0.63\\
 Caida & 2,052 & 0.97 & 206 & 0.97 & \phantom{0}1 & 0.51 \\
 \hline
\end{tabular}
\caption{Example (rounded) model sizes and accuracies per dataset.}
\label{tab:model-summary}
\vspace{-2.5em}
\end{table} 
}

Since the \textit{decision trees} exhibit competitive accuracy
while taking moderately low space when compared to
the other models (\cref{tab:model-summary}), the learned filters in the false
positive rate experiments in \cref{SEC:ANALYSIS} use \textit{decision tree} models. For construction
and query time experiments, we still compare all model variants of the learned
filters.

\para{Vectorization} To employ the models, we adapted our feature vectorization methods to the dataset. 
To train the Malicious URL and Shalla models, $20$ URL features such as hostname length, character counts, 
and number of directories were vectorized using an approach based on \cite{features:url}. 
For the Ember model, we used the vectorized features defined by the benchmark~\cite{2018arXiv180404637A}. 
For the CAIDA model, we vectorized the source and destination IP addresses and trained models to predict 
whether a network trace followed the TCP or UDP protocol.

\para{Training set proportion} 
We run a short experiment increasing the proportion of negative keys included in the training set 
and record the resulting FPR of the learned filters on $10$ million uniformly random queries on the original dataset. 
Across trials, the training set consists of all positive keys and a varying number of negative keys. 
For each of the three trials, we set the learned filter size to match an \textsc{AdaptiveQF} with $q=\lceil\log_2(n)\rceil$ and $r=5$. 
\Cref{fig:model-degrad} demonstrates that across trials, the median FPRs for all filters and models stayed 
relatively consistent with only a slight negative trend with the URL dataset, 
so we determine a train set proportion of $30\%$ negative keys to be sufficient for later experiments.

\textbf{When using the whole positive key set for training a model, the proportion of negative keys 
included in the training set can have low correlation with the quality of learned filter predictions on the overall dataset.}

\para{Model space proportion} 
Later experiments fix the parameters of the learned model while changing the size of the backup filter. 
We run a short experiment considering the opposite direction: how changing the model size affects 
performance when the overall filter size is fixed. 
We perform this experiment by fixing the learned filters to use large Random Forest Classifiers that gradually increase in size.

Across trials, we increase the number of leaf nodes used in the Random Forest Classifier. 
We fix the overall size of the learned filter to be larger than the model with the most 
leaf nodes and use the same size for each filter within the same dataset 
(described further in \cref{tab:model-proportion-parameters}). 
Any remaining space in the total filter size after building the model is allocated to the backup filters. 
After inserting the positive keys into the filter, we record the resulting false positive rates after 
performing $10$ million uniformly random queries on each dataset.

\begin{table}[h]
\small
\begin{tabularx}{\columnwidth}{|X|X|X|X|X|}
 \hline
 Dataset Name & Estimator Count & Min Leaf Nodes & Max Leaf Nodes & Filter Size (Bytes)\\
 \hline
 URL        & \phantom{0}30    & \phantom{0}5     & \phantom{0}30    & \phantom{0,}500,000 \\
 Ember      & \phantom{0}80    & \phantom{0}2     & \phantom{0}10    & \phantom{0,0}70,000 \\
 Shalla     & 150   & 20    & 140   & 4,000,000 \\
 Caida      & 120   & 20    & 100   & 3,000,000 \\
 \hline
\end{tabularx}
 \caption{Parameters used in model proportion experiments.}
\label{tab:model-proportion-parameters}
\vspace{-2.5em}
\end{table}

As anticipated from \cref{thm:space}, \cref{fig:changing-model} demonstrates that across all datasets, 
decreasing the proportion of space dedicated to backup filters eventually leads to an increase in 
false positive rate because the learned filter becomes overly dependent on model predictions. 
With less allotted space, the backup filters become more prone to reporting false positives due to an 
increased likelihood of hash collisions, reducing the benefits they provide for checking the model predictions.

\textbf{Learned filters need sufficient space budget for backup filters to offset false positives generated 
by model predictions--if a larger, complex model is needed, then the overall false positive rate of the learned filter may \textit{increase}.}

\subsection{Query Generation}
To simulate alternative filter uses, we consider four possible types of
distributions which the query workload $Q$ follows. 

\para{One-pass} In the one-pass test, each element in the dataset is queried once. 
Notably, this test was used in the empirical analysis of \textsc{Ada-BF}~\cite{Dai2020} 
and Fast \textsc{PLBF}~\cite{Sato2023} in their respective papers.

\para{Uniform} For this test, an element from the given dataset is chosen uniformly at random, for a total of $10$ million queries. 

\para{Zipfian} 
To follow a Zipfian distribution, the probability of drawing an element
$x$ at index $i$ of the dataset is proportional to $1/i^z$, 
where $z$ is some constant.
Distributions following this power-law are similar to real-world practical workloads 
where a few queries are very highly repeated \cite{Marcus2023learned, Wu2024}. 
For this test, we set $z=1.5$ and randomly draw queries from the given dataset $10$ million times 
following this distribution.

\para{Adversarial} We define an \textit{adversarial query} as one drawn from a set of likely false positives. 
An "adversary" is given sufficient time to learn these candidate adversarial queries and aims to issue 
queries which increase the filter's overall FPR. 
In practice, adversaries can discover false positives through timing attacks, as positive and negative queries 
will have different latencies due to disk accesses.

For the adversarial query test, we perform tests in which the proportion of adversarial queries is 
$d \in \{0.02, 0.04, 0.06, 0.08, 0.10\}$. 
Our adversarial test is based on the workload presented by Wen et al.~\cite{Wen2025} to evaluate \textsc{AdaptiveQF}.
Within a trial with $d$ proportion of adversarial queries, we use $|Q| = 10$M uniformly random queries from the dataset. 
In the first $|Q| / 2$ queries, we record any false positives.  
For the remaining queries, every $(|Q|/ 2) /(d * |Q|)$ queries, we replace the next random query with a false positive, 
rotating through a list of false positives. 
We record the FPR after performing all queries, including adversarial ones.

During the adversarial test, for $n$ filter insertions, we fix the \textsc{AdaptiveQF} to use $q=\lceil\log_2(n)\rceil$ and $r=6$, 
then match that size when constructing the learned filters using parameters from \cref{tab:parameter-summary}.

\subsection{Experiment categories}
\label{sec:experiment-categories}

\para{Varied query distributions} 
For this set of experiments, within a trial, we choose a dataset 
and construct a query set following one of the distributions (one-pass, uniform, Zipfian, adversarial), 
using the \textit{same} query set for all filters within a dataset. 
Then, we construct the filters according to \cref{sec:filter-const} and insert all the positive keys into each filter. 

We have all filters evaluate the complete query set and record the resulting FPR. 
We perform three trials, where in each trial we rebuild all filters (including retraining models), 
but the query set is not changed for the dataset.
When performing the adversarial test, we chose not to use the adaptive variant of the stacked filter because it
requires as input the distribution of the negative queries in the workload,
which will change according to the false positives which the adversary detects.

\para{Dynamic workload} For this experiment, we emulate the setting where the contents of the filter 
change in distribution while queries are performed. We first construct an \textsc{AdaptiveQF}, 
fixing it to use $q=\lceil\log_2(n)\rceil$ and $r=5$ for datasets with $n$ positive keys. 
We construct an \textsc{Ada-BF} and \textsc{PLBF} to each use the same space after training models 
according to the parameters described in \cref{tab:parameter-summary}. 
We also construct a stacked filter matching the space usage.

We start by defining a set of $10$ million uniformly distributed queries from a dataset. 
Then, we begin querying the filters, reporting the instantaneous FPR whenever $1\%$ of the queries have been finished. 
The \textit{instantaneous false positive rate} is the FPR of a filter on all queries. 
The \textsc{AdaptiveQF} does not perform adaptations during this check.

Whenever $10\%$ of queries are finished, we perform a churn of the dataset. 
We set aside a replacement set: a subset of the negative keys from the dataset equivalent in size to the set of positive keys. 
This size requirement ensures no negative keys are inserted as duplicates 
(we leave out the Shalla dataset since it has more positive than negative keys). 
There are $10$ churns total, each with replacement size $n/5$. 
On the $j$-th churn, we swap a portion of the inserted keys and replacement set, 
starting with the keys in position $start=n/5*(j\%5)$ and ending at position $start + n/5$.
Overall, each churn replaces $20\%$ of the inserted key set with negative keys, 
so while performing queries the insertion set is gradually completely replaced with non-keys 
before being returned to the original insertion set.

In response to churns, we rebuild the filters so that they have the correct contents.
The learned filters' internal models are retrained on the new positive set,
which is a common response to distribution shifts~\cite{Gama2014}. 
On each churn, the reconstructed stacked filter is given the new distribution of the dataset.

\para{Construction and query times} To evaluate the time it takes to perform different operations on each filter,
we perform separate experiments where the construction of the learned filters is adjusted. 
The original learned filter code from~\cite{Dai2020, Sato2023} trains models separately and computes all scores 
for the insertion and query sets before passing them to stored filters which are sized according to the space budget--
this process streamlines query processing for FPR analyses. We partially rewrite the filters to store the learned model. 
Now, during a query, the filters first perform model inference on a vectorized key to obtain a score and decide how to 
query its backup filter(s). 
This simulates a realistic use case in which a learned filter must respond to a query on the fly.

During construction, for each dataset, for a dataset with $n$ positive keys, 
we set $q=\lceil\log_2(n)\rceil$ for the \textsc{AdaptiveQF}, set $r=5$, 
then match the resulting size when constructing the learned and stacked filters. 
We run 3 trials of $10$ million uniformly distributed queries, recording the median construction
and amortized query times for each filter.

%% file: analysis.tex
\begin{figure*}[h]
    \centering
    \includegraphics[width=\textwidth]{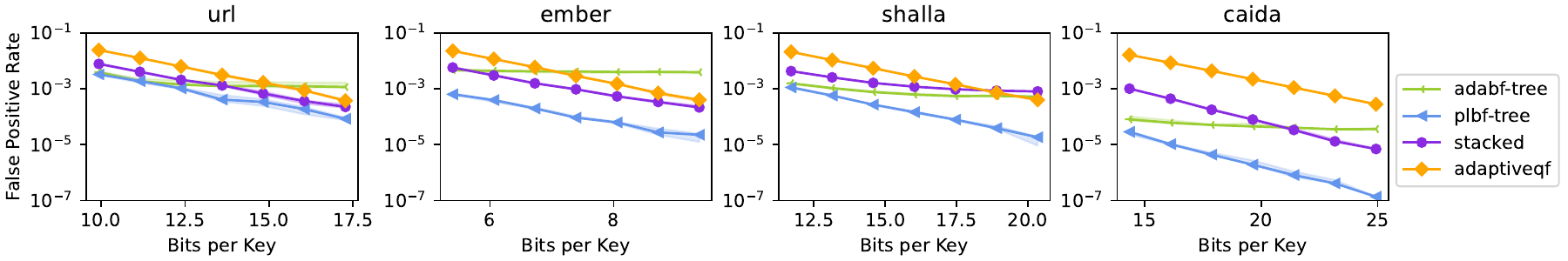}
    \caption{The FPR-space tradeoff for each dataset on a \emph{one-pass} query test.
    Learned filters perform well, with stacked filters close behind.}
    \label{fig:one-pass-results}
\end{figure*}

\begin{figure*}[h]
    \centering
    \includegraphics[width=\textwidth]{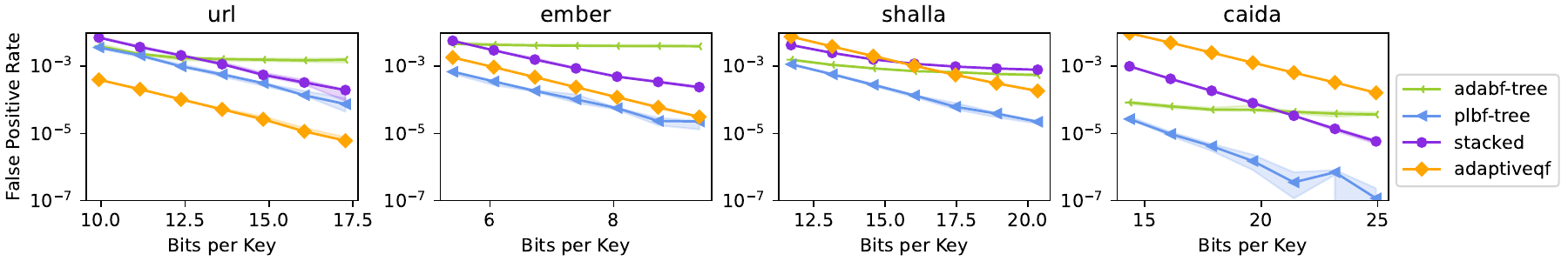}
    \caption{The FPR-space tradeoff for each dataset on 10M \emph{uniformly} distributed queries.
    Adaptive filters begin to catch up to stacked and learned filter performance.}
    \label{fig:uniform-results}
\end{figure*}

\begin{figure*}[h]
    \centering
    \includegraphics[width=\textwidth]{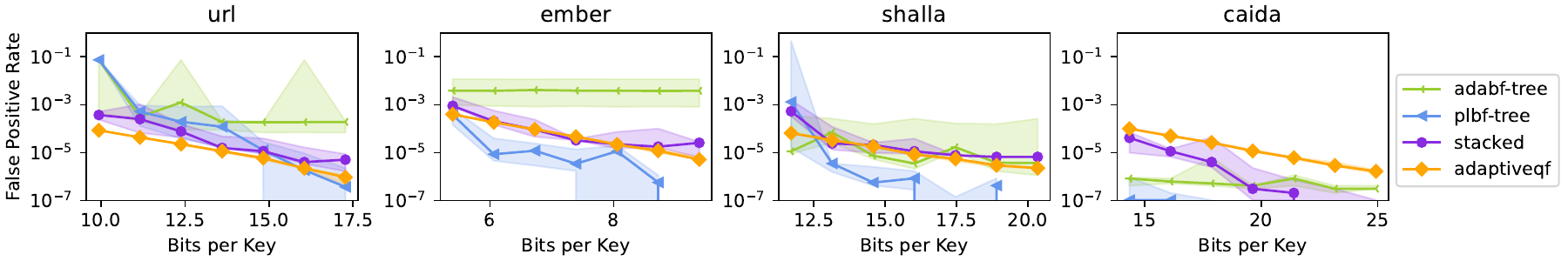}
    \caption{The FPR-space tradeoff for each dataset on 10M \emph{Zipfian} distributed queries.
    Learned filter performance becomes variable, while stacked and adaptive filters exhibit consistent and low FPRs.}
    \label{fig:zipfian-results}
\end{figure*}

\begin{figure*}[h]
    \centering
    \includegraphics[width=\textwidth]{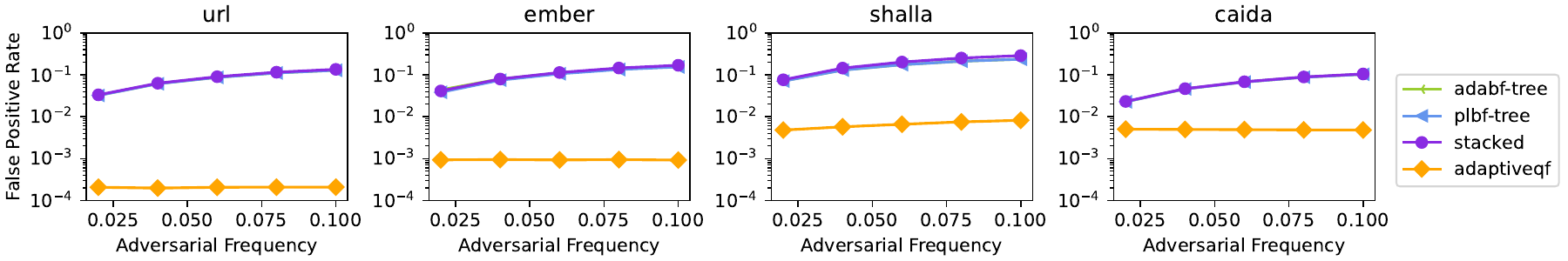}
    \caption{The effect of increasing the proportion of \emph{adversarial} queries on the FPR of fixed-size filters for each dataset.
    Adaptive filters outperform all other filter types, whose results are identical and overlap.
    }
    \label{fig:adversarial-results}
\end{figure*}

\section{Analysis}
\label{SEC:ANALYSIS}

\subsection{Varied query distributions}
\label{sec:varied query distributions}
The results for each test provide interesting insights into how the guarantees of 
learned, stacked, and adaptive filters compare. 
In the FPR-space graphs we obtain (\cref{fig:one-pass-results}, \cref{fig:uniform-results},
\cref{fig:zipfian-results}), 
we have the total filter size 
(including the trained model, if any) on the x-axis and the FPR on the y-axis. 
In general, more memory-efficient data structures will have lower curved lines. 
We plot the median FPRs across $3$ trials, then shade between the minimum and maximum values 
within trials to indicate the variability of the results.

\para{One-pass} In \cref{fig:one-pass-results},
for the learned filters, we find that the \textsc{PLBF} 
outperforms both the \textsc{AdaptiveQF} and stacked filter.
\textsc{PLBF} also performs better than the \textsc{Ada-BF}, but this is expected and was 
previously demonstrated in a comparison between the two learned filters~\cite{Sato2023}.

Since the one-pass query distribution is essentially a uniform distribution, 
the stacked filter is able to use its knowledge of the distribution and reduce its false positive
rate through layers of filters. 
Thus, it performs better than the \textsc{AdaptiveQF}, though it is still outperformed 
by the \textsc{PLBF}.

With the \textsc{AdaptiveQF}, since each query is only performed once, it is unable to take advantage of 
multiple corrected false positives.
So, in this workload, the \textsc{AdaptiveQF} behaves similarly to a traditional quotient filter,
corresponding to its relatively higher FPRs compared to the other filter types.

\para{Uniform} 
With uniformly distributed queries, we find in \cref{fig:uniform-results} 
that with the learned filters, the reported false positive rates remain very similar to the results
from the one-pass test, since the uniform query distribution matches the training set distribution,
allowing the model results to be generalizable.

The stacked filter performance also remains consistent with its one-pass results, 
since the distribution shape of the sampled negative values is similar to that of the 
uniformly-distributed query workload.

Regarding the \textsc{AdaptiveQF}, the gap between its performance and that of the other filters has become closer.
This is likely due to the \textsc{AdaptiveQF} now being able to take advantage of its false positive
adaptations, so if any false positive queries are repeated the \textsc{AdaptiveQF} later correctly 
classifies them.
In a uniform workload with a larger number of queries than unique elements, a false positive is more likely
to repeat, allowing the \textsc{AdaptiveQF} more opportunities
to benefit from its prior adaptations.
In particular, since the URL dataset has the lowest number of unique negative values,
its uniform workload contains the highest number of repeated negative queries across the datasets,
resulting in the \textsc{AdaptiveQF} performing the best on the dataset.

\begin{figure*}[h]
    \centering
    \includegraphics[width=\linewidth]{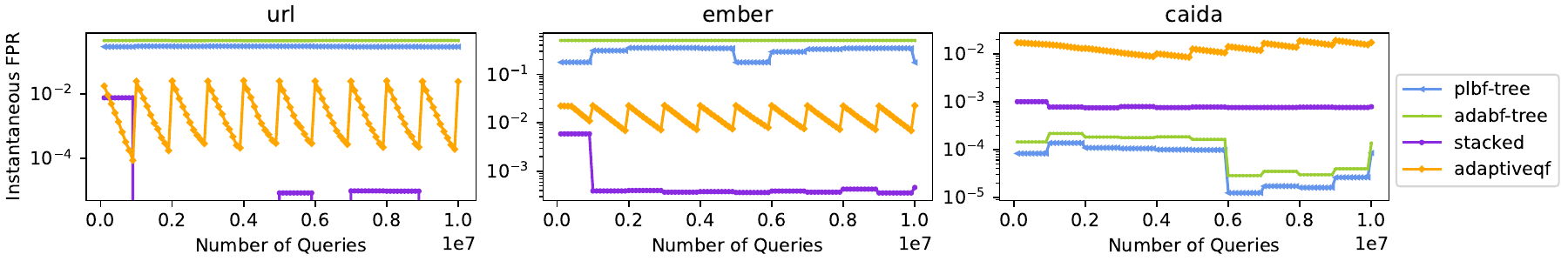}
    \caption{Instantaneous FPR of filters in response to \emph{dynamic} workload \textit{with model retraining}.
    Even with retraining, adaptive and stacked filters can still exhibit lower FPRs.}
    \label{fig:dynamic-rebuild}
    \vspace{-1.5em}
\end{figure*}

\para{Zipfian} 
Results in \cref{fig:zipfian-results} 
highlight the significance of filters' 
theoretical assumptions on the distribution of the insertion and query sets. 

For all datasets with an Zipfian query load, 
the learned filters have significantly varied FPRs, 
with minimum and maximum FPRs
for the same filter variant potentially being orders of magnitude apart.
The learned filters have particularly poor results in the Shalla dataset, 
with a maximum FPR close to $50\%$, but also reach near-zero FPRs on the other datasets.
There are two main reasons why the learned filter may exhibit such variance on these Zipfian query workloads.

Firstly, the model may be making inaccurate predictions after training. 
Between trials, the model and filter were reconstructed. 
During training, the model may fit to data points which appear infrequently in the workload 
at the cost of lower accuracy on frequently queried elements.
If the model predicts a higher score for a negative key, 
then it the key is more likely to be returned as a false positive. 

Secondly, the learned filters are inherently unable to adapt.
If the model predicts a false positive, it repeats this mistake when the query is repeated.
If the model predicts a negative but the backup filter returns a false positive, 
it cannot make changes to the backup filter to adapt to the result. 
So, in the case of the Zipfian workload, increasing the repetitions of a query can have 
drastic effects on the filter's overall FPR if both the model and backup filters 
repeat mistakes which cannot be corrected, contributing to inconsistent results across trials.
In the other direction, if correct model decisions are frequently repeated, the learned filter
can achieve extremely low FPRs, such as in the Caida dataset.

The stacked filter uses its knowledge of the workload distribution well.
For a Zipfian workload, the query sample that the filter used for optimization likely contained
the set of highly repeated negative queries, allowing the stacked filter to store those values
and greatly decrease the odds of them becoming false positives.
So, the stacked filter demonstrates performance competitive with the \textsc{AdaptiveQF}, 
even outperforming
it on the Caida dataset.

The \textsc{AdaptiveQF} also has a consistently low FPR. 
In a Zipfian query workload, if a false positive occurs, it is likely to be highly repeated.
Out of all static query distributions, the Zipfian workload provides the greatest
opportunity for the \textsc{AdaptiveQF} to benefit from adaptations,
as demonstrated by its greatly improved performance compared to the one-pass and uniform workloads. 

\para{Adversarial} For the adversarial test (\cref{fig:adversarial-results}), 
as the proportion of adversarial queries increases, the learnedf ilters demonstrate increasing FPR.
Again, since the learned filters cannot adapt, repeating previous false positives results in an increased overall FPR.
Overall, since learned filter FPRs are highly dependent on the model quality for a particular dataset, 
the benefits from utilizing learned predictions are inconsistent compared to the strong adaptability guarantees of adaptive filters. 
We confirm that, as described in \cite{Mitzenmacher2018}, learned filters are competitive when the  query distribution is similar to 
the distribution that the filters are trained on. 

The same drop in performance is present for the stacked filter: as the negative set changes, the stacked filter cannot update its contents,
resulting in increasing FPRs which are drastically higher compared to the \textsc{AdaptiveQF}.
Likewise, stacked filters rely on the assumption that the query distribution remains relatively consistent, but introducing more negative queries
makes the performance of the stacked filters less consistent.

For the \textsc{AdaptiveQF}, its FPR remains generally constant and low across all datasets with the adversarial workload.
Adaptive filters are able to consistently alleviate errors caused by repeating a particular query in an arbitrary workload.

\textbf{When query patterns include repeated (skewed) queries, adaptive filters often exhibit lower FPRs and 
more consistent negative correlations between space budget and FPR compared to learned and stacked filters. 
Learned filters are more appropriate for tasks where the query distribution closely matches the dataset distribution (such as one-pass classification),
while stacked filters are appropriate when query workload distributions are not expected to change significantly.}

\subsection{Dynamic workload}

\begin{figure*}\includegraphics[width=.99\linewidth]{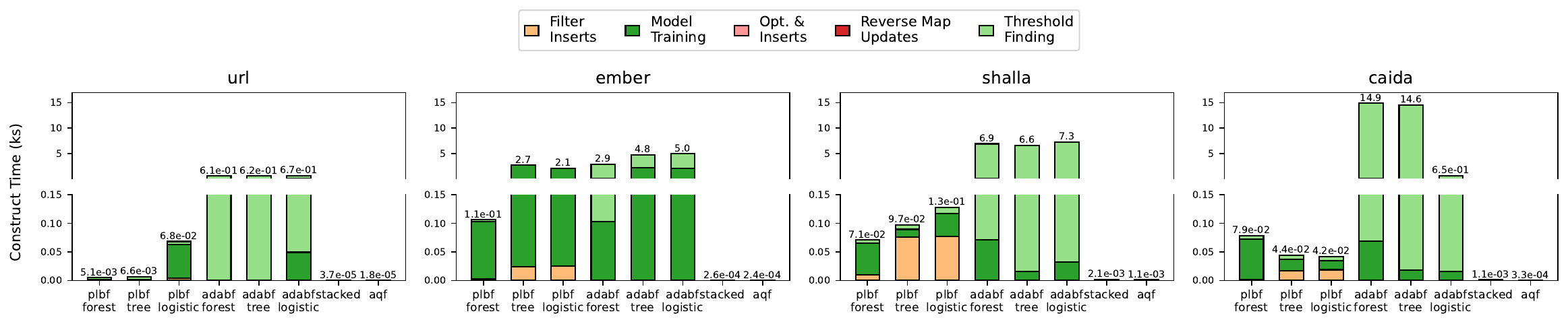}
    \caption{Overall filter \emph{median construction} time on each dataset. Model training time dominates learned filter construction,
    resulting in stacked and adaptive filters demonstrating orders of magnitude faster construction.}
    \label{fig:overall-construct}
\end{figure*}

\Cref{fig:dynamic-rebuild} indicates that retraining the model in response to
dataset distribution shifts may improve FPR, but these effects are
inconsistent. For instance, when the positive key set is replaced, the learned
filters experience massive improvements in FPR on the Caida dataset, but on the
URL and Ember datasets their FPR generally increased. With distribution shifts,
the correlation between key features and presence in the set can weaken, making
model predictions unreliable. However, even when models make good predictions,
overall FPRs are still limited by the learned filters' restriction to repeat
the same predictions, resulting in static FPRs between churns. 
Additionally, \cref{fig:overall-construct} shows that the
time needed to train a model is costly and can bottleneck the overall
construction time of a filter, making this retraining procedure less practical.

For the stacked filters, their performance in 
\cref{fig:dynamic-rebuild} demonstrates that
under conditions where periodic updates to the dataset are expected, simply relearning the distribution of negatives
corresponds to highly competitive and consistent performance. 

The robustness of the \textsc{AdaptiveQF} is demonstrated by \cref{fig:dynamic-rebuild}. 
At every churn, the instantaneous FPR of spikes before rapidly decreasing due to adaptations, 
maintaining a relatively low overall FPR.
The \textsc{AdaptiveQF} demonstrates that even with low FPRs, it can still adapt to achieve even lower FPRs over the course of
a set of queries, while the learned filters remained fixed at particular FPRs between churns.
However, in this workload, the \textsc{AdaptiveQF} does not adapt to enough false positives between churns to achieve 
instantaneous FPRs lower than the stacked filters.

\textbf{
Incorporating a machine learning model into a filter greatly limits its robustness. 
Models do not generalize well across dynamic workloads, and retraining does not guarantee 
improved performance on new distributions. The ability of a data structure to adapt to the
query workload is crucial in these settings.}

\begin{figure*}\includegraphics[width=.99\linewidth]{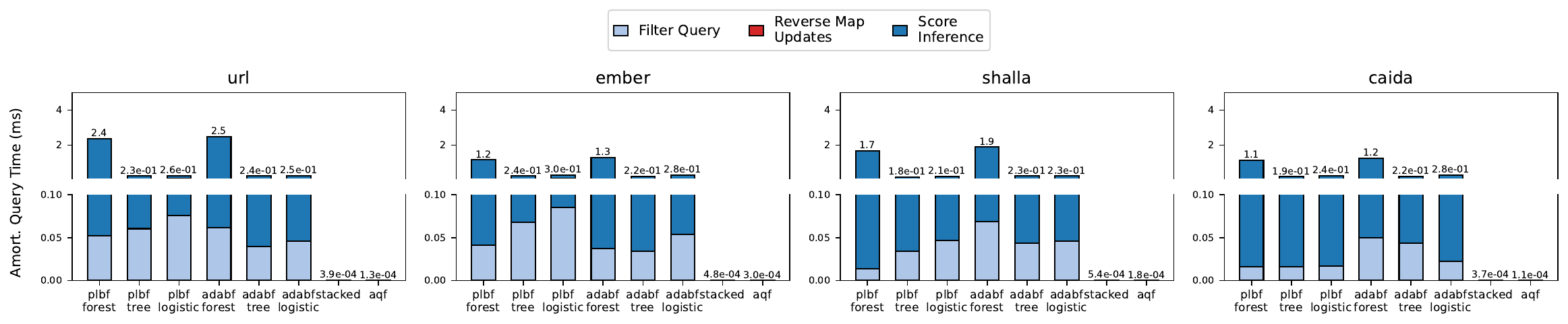}
    \caption{Overall filter \emph{amortized query} time on each dataset. The cost of querying internal models for scores is much higher
    than that of querying backup filters, corresponding to stacked and adaptive filters exhibiting orders of magnitude faster queries.}
    \label{fig:overall-query}
\end{figure*}

\subsection{Construction and query times}
\para{Construction time} 
The construction times (in kiloseconds) for each filter and the proportion of time 
spent on different operations are plotted in \cref{fig:overall-construct}. 
The construction operations were divided into \textit{Filter Inserts} 
(constructing the traditional filter structure and inserting elements into it), 
\textit{Model Training} (constructing and training a machine learning model on a dataset), 
\textit{Threshold Finding} (querying the model for scores and deciding how to use these scores to assign groups), 
and \textit{Reverse Map Updates} (inserting elements into an auxiliary reverse map for the \textsc{AdaptiveQF}).

In all datasets, the construction of the \textsc{Ada-BF} is slowest largely due to the time 
it takes to find optimal score thresholds for group assignments. 
Note that future work could reduce this overhead--currently, the \textsc{Ada-BF} implementation~\cite{Dai2020} 
loops over and tests a range of thresholds. 
The \textsc{Fast PLBF} implementation~\cite{Sato2023} is an example of work which optimizes 
the construction of thresholds for prior learned filters~\cite{Vaidya2021}.

In~\cref{fig:overall-construct}, we find that of the learned filter variants,
those based on the \textsc{PLBF} are the fastest to construct. However, across
all the datasets it is still $100\times$ slower to construct than the
\textsc{AdaptiveQF} and stacked filter \textit{at minimum}. This occurs because
the construction time of the \textsc{PLBF} is dependent on its model training
time. Even when score thresholds are efficiently established and backup filter
construction is quick, model training time can take longer than the entire
\textsc{AdaptiveQF} construction. This observation indicates that model
training and construction is a significant overhead.
Meanwhile, non-learned adaptive filters like the \textsc{AdaptiveQF} only
consider the number of insertions (and potential adaptations for initial
collisions) during construction time. The stacked filter takes some time to
optimize based on the given negatives, but this process is still much faster
than model training.

\textbf{Learned filter construction time is highly dependent on the time it
takes to train the model. In settings where more complex models are needed to
capture correlations between features and classifications, it becomes less
practical to rebuild learned filters in the face of new query distributions.}

\para{Query Time} 
The amortized query times (in milliseconds) and the proportion of time spent on 
different operations after performing experiments as described
in \cref{sec:experiment-categories} are plotted in \cref{fig:overall-query}. 
The included query operations were the \textit{Filter Query} (probing the traditional filter(s) for an element), 
\textit{Score Inference} (having an ML model generate a score for an element), 
and \textit{Reverse Map Updates} (updating the \textsc{AdaptiveQF} using auxiliary data in response 
to false positives).

The most striking observation is that all amortized query times for both
learned filters are up to $10^4\times$ slower than those of the
\textsc{AdaptiveQF} and stacked filter. 
Across all datasets, learned filter query latency is dominated by the time spent
performing model inference, even when the model is much smaller (in
the cases of variants using decision trees or logistic regression models).
This weakness contributes to learned filter query times \textit{at least} $10^3\times$
slower than those of the \textsc{AdaptiveQF}. Notably, the vectorization of key
features was implemented as a pre-processing step, and could serve as an
additional overhead for on-the-fly query processing. 

Meanwhile, the \textsc{AdaptiveQF} and stacked filter only depend on traditional filter
operations and are extremely quick to process queries.
However, due to stacked filters consisting of multiple layers of individual filters,
the computational costs compound and result in query times up to $3\times$ slower than
adaptive filters.

\textbf{The computational cost of model inference in learned filters is orders
of magnitude more expensive than traditional filter operations even when
simpler learned models are used, making learned filters impractical in settings
where query throughput matters.}

%% file: discussion.tex
\section{Discussion}

Filters have been a cornerstone of systems design for over five decades, with
recent paradigm-shifting advancements 
(learned, stacked, and adaptive filters) 
promising orders-of-magnitude improvements in FPRs.
This paper provides the first unified evaluation, revealing that the choice of
filter paradigm depends critically on application characteristics that prior
work often overlooks. Our evaluation yields three principal insights:

(1) First, \emph{learned filters
are not a universal improvement over traditional filters}. 
While they achieve
large reductions in false positive rate (up to $10^2\times$ better) when query
distributions match training data, this advantage can degrade under distribution shift, 
adversarial workloads, or repeated queries where ML model errors are amplified.

(2) Second, \emph{computational overhead fundamentally changes the calculus}.
Learned filters' model inference costs (up to $10^4\times$ slower than
traditional filter operations) have been systematically overlooked in prior
evaluations. 
For applications where query throughput matters, this overhead can
negate any space-efficiency gains. Adaptive and stacked filters, by contrast,
maintain predictable and efficient query performance while still achieving competitive FPRs.

(3) Third, \emph{robustness and generalizability vary across paradigms}. Adaptive
filters provide the strongest and most robust guarantees: bounded false
positive rates for arbitrary query sequences without requiring assumptions
about workload distributions. Stacked filters offer excellent performance when
query distributions are known and stable, achieving up to $10^3\times$ lower
false positive rates on skewed workloads. Learned filters, despite their
sophistication, tend to be least robust, 
exhibiting higher variance and potentially
unpredictable behavior 
outside their training regime.

\introparagraph{Future work: robust hybrids} 
Our study raises a critical question: can hybrid approaches combine the
robustness of adaptive filters with the distribution-awareness of learned or
stacked filters? Our evaluation framework and findings provide a foundation for
exploring these directions
by emphasizing considerations that should not be overlooked in filter evaluation
and characterizing the performance of modern filter paradigms.
We leave these potential designs for future work.

%% file: ack.tex
\begin{acks}
This work is supported by the NSF grants 2517201, 2513656, and DGE-2439018. 
We thank Yuvaraj Chesetti and Aidan Ih for providing support for the implementation of the experiments.
\end{acks}

%% file: appendix.tex
\section{Nomenclature}
\begin{table}[h]
\begin{tabularx}{\columnwidth}{lX}
 \toprule
 Symbol & Definition  \\
 \midrule
 FPR & (empirical) false positive rate  \\
 $s$ & Model-output confidence score describing likelihood of key being present in set  \\
 $t$ & Score threshold defining when learned filters query backup filters   \\
 $b$ & Bits used per key in a learned filter's backup filter \\
 $\alpha$ & The load factor of a filter \\
 $\zeta$ & The number of bits used by a learned filter's internal ML model \\
 $f_m$ & False positive rate of a learned filter's ML model \\
 $f_n$ & False negative rate of a learned filter's ML model \\
 $f_b$ & FPR of a learned filter's backup filter(s) \\
 $k$ & Number of score-dependent groups keys are distributed across in a learned filter \\
 $c$ & How many times more non-keys each score-divided group of keys is expected to contain compared to the next group, used by the \textsc{Ada-BF} \\
 $N$ & Number of segments the score space is split into before assigning keys to groups, used by the \textsc{PLBF} \\
 $\epsilon$ & Theoretical false positive rate for a filter \\
 $x$ & Example queried item \\
 $S$ & Set of keys represented by a filter \\
 $h(x)$ & Hash of element $x$ \\
 $h_0(x)$ & Higher-order bits of $h(x)$ comprising the quotient \\
 $h_1(x)$ & Lower-order bits of $h(x)$ comprising the remainder \\
 $p$ & Length of fingerprint stored in \textsc{AdaptiveQF} \\
 $q$ & Length of hash quotient used by \textsc{AdaptiveQF} \\
 $r$ & Length of hash remainder and (future extensions) used by \textsc{AdaptiveQF} \\
 $n$ & Number of unique positive key elements inserted in a filter\\
 $i$ & Index of element in some permutation of its dataset \\
 $j$ & Current number of churns performed in the dynamic query workload \\
 $z$ & Zipfian constant describing power-law relation of randomly drawn elements \\
 $Q$ & Multiset of elements in the query workload \\
 $Q_N$ & Multiset of negative elements in the query workload \\
 $Q_{FP}$ & Multiset of false positives reported from query workload \\
 \bottomrule
\end{tabularx}
\label{tab:nomenclature}
\vspace{-1.5em}
\end{table} 

%% file: filters.tex
\section{Filters}
\label{SEC:FILTERS}

Filters, such as Bloom~\cite{Bloom1970}, quotient~\cite{Pandey2017, Bender2012,
Dillinger2009, Einziger2016, Dayan2023, Pandey2021}, and cuckoo~\cite{Fan2014,
Breslow2018} filters, maintain an approximate representation of a set. The
approximate representation saves space by allowing queries to occasionally
return a false positive. 
For a given false positive rate (FPR) $\epsilon$, a membership query to a
filter for set $S$ is guaranteed to return \texttt{present} for any $x \in S$, 
while returning \texttt{present} with probability at least $\epsilon$ for any $x \notin S$.
For $|S| = n$, space depends on $(n, \epsilon)$ and is typically far smaller
than storing $S$ explicitly.

\subsection{Types of filters}
Filters are commonly classified as \emph{static}, \emph{semi-dynamic}, or \emph{dynamic}.
Static filters (e.g., XOR~\cite{Graf2020}, Ribbon~\cite{Dietzfelbinger2026} filters) 
require the complete item set before
construction. 
Semi-dynamic filters like Bloom~\cite{Bloom1970} and Prefix
filters~\cite{Even2022} support insertions but typically require an estimate of
the final set size to meet a target FPR, which complicates deployments where
$|S|$ is unknown.
Dynamic filters (quotient~\cite{Bender2012, Pandey2017, Dillinger2009,
Pagh2005, Einziger2016, Dayan2023} and cuckoo~\cite{Fan2014,
Breslow2018} filters) support both insertions and deletions and handle unknown sets and
sizes, making them increasingly popular for real-world applications.
Dynamic filters require $n\log\varepsilon(1/\epsilon) + \Omega(n)$ bits: quotient
filters use $n\log(1/\epsilon) + 2.125n$ bits~\cite{Pandey2017} and cuckoo
filters use $n\log(1/\epsilon) + 3n$ bits~\cite{Fan2014}. 
Bloom filters use about $1.44 \cdot n \cdot \log(1/\varepsilon)$ bits, but
dynamic filters can approach 
$n \cdot \log(1/\varepsilon) + \Omega(n)$ bits; 
Bloom filters are only smaller at very large $\varepsilon$ (high FPR).

\subsection{Fingerprint-based filters}
State-of-the-art dynamic filters are fingerprint-based. 
They compactly store short, lossy fingerprints in a table using \defn{quotienting}~\cite{Pagh2005}. 
In quotienting, a $p$-bit fingerprint $h(x)$ is divided into two parts: 
the first $q$ bits  $h_0(x)$ is called the \defn{quotient} and the remaining $r=p-q$ bits $h_1(x)$ is called the \defn{remainder}. 
The quotient is stored implicitly and only the remainder is stored explicitly to save space.

Quotient filters~\cite{Pandey2017, Bender2012, Einziger2016, Dillinger2009, Pandey2021} use Robin Hood hashing (a variant of linear probing) 
to store the remainders in a linear table. 
To resolve soft collisions (i.e., when two fingerprints share the same quotient but have different remainders) it uses 2-3 extra metadata bits
to resolve collisions and efficiently perform queries.

\section{Modern filter paradigms}
Recent advancements in filter design leverage dataset features,
workload knowledge, or query feedback to improve false positive rates
asymptotically.
We provide short summaries of each of these new filter paradigms:

\introparagraph{Learned Filter} \emph{During initialization},
inserts keys into backup filter(s), then trains an internal ML model using
features of those inserted keys.
\emph{During deployment}, the model generates a score describing the likelihood
of a queried element being an inserted key, which is used to smartly decide
how to query the structure's backup filters.

\introparagraph{Stacked Filter} \emph{During initialization},
a sample of the query workload is used to build layers of cascading filters,
where the first large layer stores all positive keys, then the following layers
alternate between storing all negative or all positive values which were false
positives in the previous filter layer.
\emph{During deployment}, a queried element traverses the layers sequentially.
If a layer returns negative, the element is returned as a (potentially false)
positive if the layer stored negatives and returned as a (true) negative if the layer
stored positives.
If the element makes it through all layers of filters, the final filter determines the classification
of the queried element.

\introparagraph{Adaptive (Quotient) Filter} \emph{During initialization},
in addition to storing all elements in a quotient filter, a (disk-resident) reverse map
is maintained which associates each partial hash of an inserted element with the complete original key.
\emph{During deployment},l upon receiving feedback about a false positive result,
the filter consults the reverse map to obtain sufficient information to update
the structure and ensure the false positive is not repeated in the future.

%% file: app-learned.tex
\section{Learned filters}
\subsection{Base (sandwiched) learned filters}
In the formal learned filter proposal \cite{Mitzenmacher2018},
learned filters simply depended on a model score threshold to decide
whether or not the the ML model's prediction was a (potentially false)
positive or a negative result which required verification from a backup filter.
The sandwiched filter was proposed as an extension of this design,
where an additional preliminary filter is checked before queries
continue to the trained model.
The preliminary filter ensures that
(potentially many) true negatives cannot be misclassified
by the ML model.
The (sandwiched) learned filter structure is represented
by \cref{fig:learnedfilter}.
In their experimental evaluations, the \textsc{Ada-BF}, \textsc{PLBF},
and Stacked filters all demonstrated that their implementations provided
better FPRs than the (sandwiched) learned filter \cite{Dai2020,Sato2023,Deeds2020}.
Notably, it was found that in practical workloads, to optimize for space the sandwiched
learned filter reduces to the base learned filter \cite{Dai2020}.
\iflongversion
\begin{figure}[h]
\includegraphics[width=0.9\linewidth]{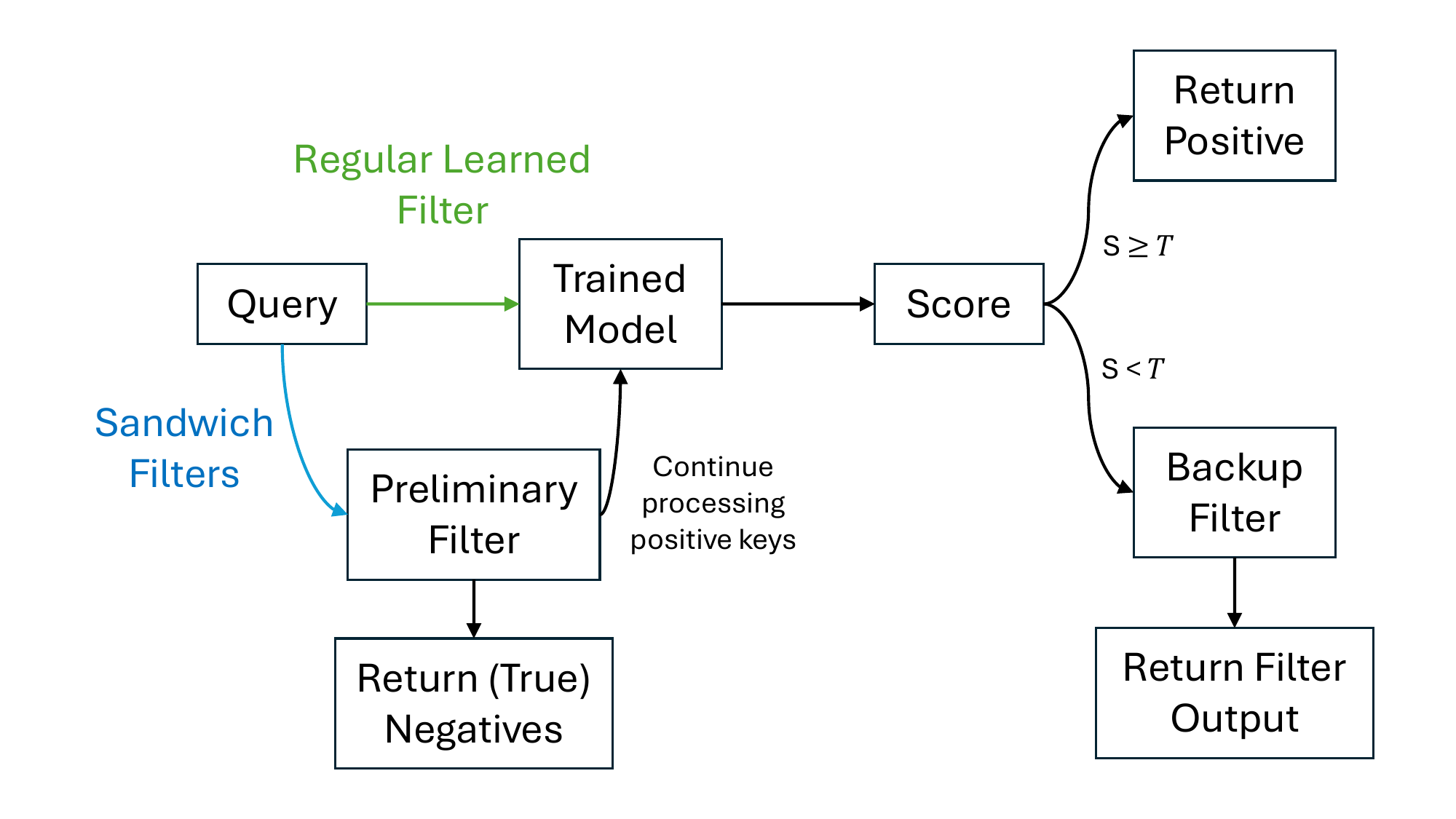}
\caption{(Sandwiched) learned filter structure. 
A model prefilters positive keys before testing a backup filter for negative candidates. 
For the sandwich filter, an additional traditional filter is queried to remove true negatives
before continuing to the trained model.}
    \label{fig:learnedfilter}
    \vspace{-1.5em}
\end{figure}
\fi

\subsection{Learned filter distribution assumptions}
Learned filters depend on the model training set closely matching
the distribution of the query workload~\cite{Mitzenmacher2018}.

\Cref{thm:fpr} formally describes this finding: 
\begin{theorem}{\cite{Mitzenmacher2018}}
\begin{samepage}
    For some learned Bloom filter, consider a test set T, and a query set Q, where T and Q are both sampled from some distribution. 
If x is the empirical false positive rate of the filter on T and y is its empirical false positive rate on Q, then
    $$Pr(|x - y| \ge \delta) \le e^{-\Omega(\delta^2 min(|T|,|Q|))}$$
\end{samepage}
\label{thm:fpr}
\vspace{-1.5em}
\end{theorem}

\Cref{fig:zipfian-results,fig:adversarial-results,fig:dynamic-rebuild}
describe experiments where the query workload distribution differs greatly
from the training set distribution.
In these cases, the learned filters demonstrated variable or worsened
FPRs, affirming the insights from \cref{thm:fpr}.

\subsection{Model training results}
\Cref{fig:score-dist-forest,fig:score-dist-tree,fig:score-dist-logistic} 
demonstrate examples of the positive and
negative key distributions across each dataset after one trial of model
training using random forests, decision trees, and logistic regression models,
respectively. The scores describe the model's predicted likelihood that the
given (vectorized) key is a positive key. Notably, when the model becomes
extremely small in the case of the logistic regression models
(\cref{tab:model-summary}) in \cref{fig:score-dist-logistic}, the model
predictions become unreliable, with positive and negative key scores
demonstrating high overlap.

\iflongversion
\begin{figure*}[h]
    \centering
    \includegraphics[width=\textwidth]{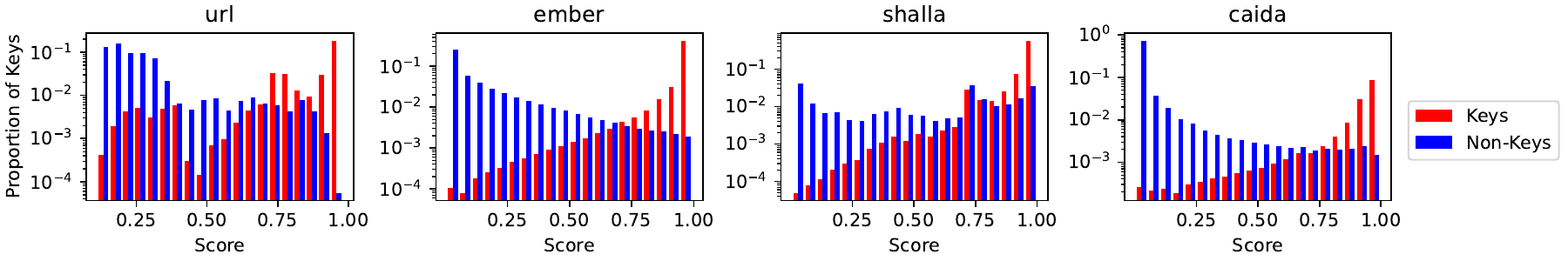}
    \caption{The distribution of \emph{Random Forest} model scores for each dataset. 
    Negative keys tend to skew towards lower scores, while positive keys skew towards higher scores.}
    \label{fig:score-dist-forest}
\end{figure*}
\begin{figure*}[h]
    \centering
    \includegraphics[width=\textwidth]{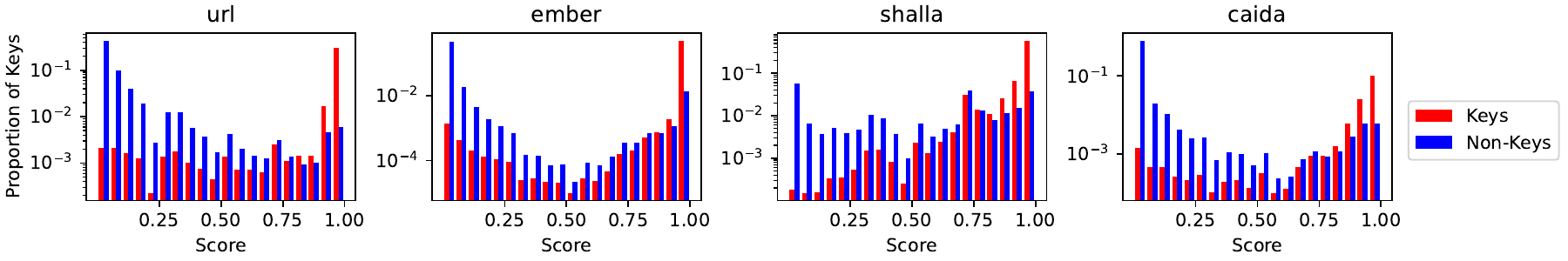}
    \caption{The distribution of \emph{Decision Tree} model scores for each dataset.
    Positive keys still tend to skew towards higher scores, but negative keys have less consistent skew across datasets.}
    \label{fig:score-dist-tree}
\end{figure*}
\begin{figure*}[h]
    \centering
    \includegraphics[width=\textwidth]{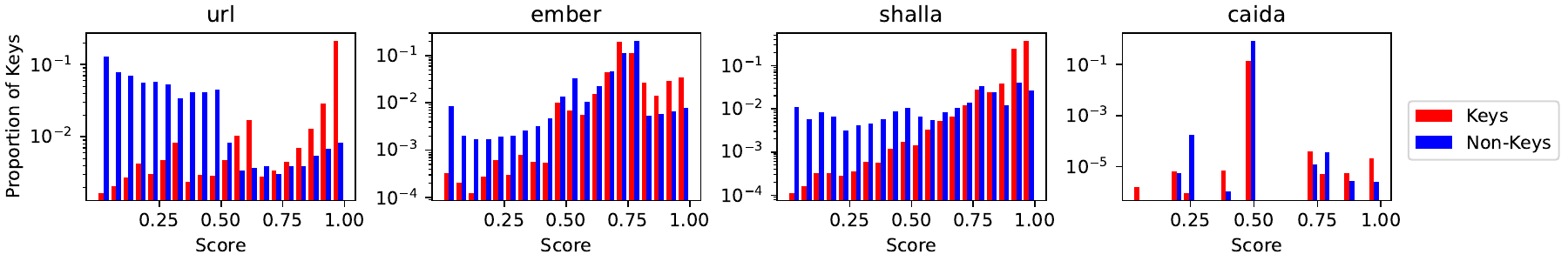}
    \caption{The distribution of \emph{Logistic Regression} model scores for each dataset.
    Negative key scores have higher overlaps with positive key patterns.}
    \label{fig:score-dist-logistic}
\end{figure*}
\fi

%% file: app-stacked.tex
\section{Stacked Filters}
\Cref{thm:stacked}~\cite{Deeds2020} formalizes the FPR and
space guarantees of the stacked filter, stating that the overall false positive
rate and space usage of the filter is dependent on the configuration of the
filter layers comprising the overall structure.

\begin{theorem}{\cite{Deeds2020}}
\begin{samepage}
    For a given stacked filter, let $\psi$ represent the probability that a 
    negative element from a query distribution is in the set of frequently queried negatives $N_F$, 
    $P$ represent the set of positive elements,
    $l$ 
    represent the number of layers in the structure,
    $a_i$ represent the false positive rate of the $i^{th}$ layer,
    and $s(\epsilon)$ represent the size in bits for a filter layer to achieve FPR $\epsilon$.

    Then the expected false positive rate of the stacked filter is
    $$\psi \prod_{i=0}^{(l-1)/2}a_{2i+1} + (1-\psi)( \prod_{i=1}^{l} a_i + \sum_{i=1}^{(l-1)/2} (\prod_{j=1}^{2i-1} a_j)(1-a_{2i})  )$$

    and the overall size of $S$ is given by
    $$\sum_{i=0}^{(l-1)/2}s(a_{2i+1})*(\prod_{j=0}^{a_2j})+\sum_{i=1}^{(l-1)/2}s(a_{2i})*\frac{|N_F|}{|P|}*(\prod_{j=1}^{i}a_{2j-1})$$
\end{samepage}
\label{thm:stacked}
\end{theorem}

The initial layer storing all positives ends up dominating the space usage
because the FPRs of each layer ensures that exponentially fewer
elements must be stored in consecutive filter layers.

%% file: app-adaptive.tex
\section{Adaptive Filters}
An example of how the \textsc{AdaptiveQF} uses quotienting
for adaptations is depicted in~\cref{fig:aqf-ex}.
During an adaptation, when the remainder is extended it may take up additional
quotient filter slots, so the \textsc{AdaptiveQF} additionally stores
a few bits of metadata per element to keep track of which (larger) remainders
correspond to which slot.

\iflongversion
\begin{figure}[t]
    \centering
    \includegraphics[width=\linewidth]{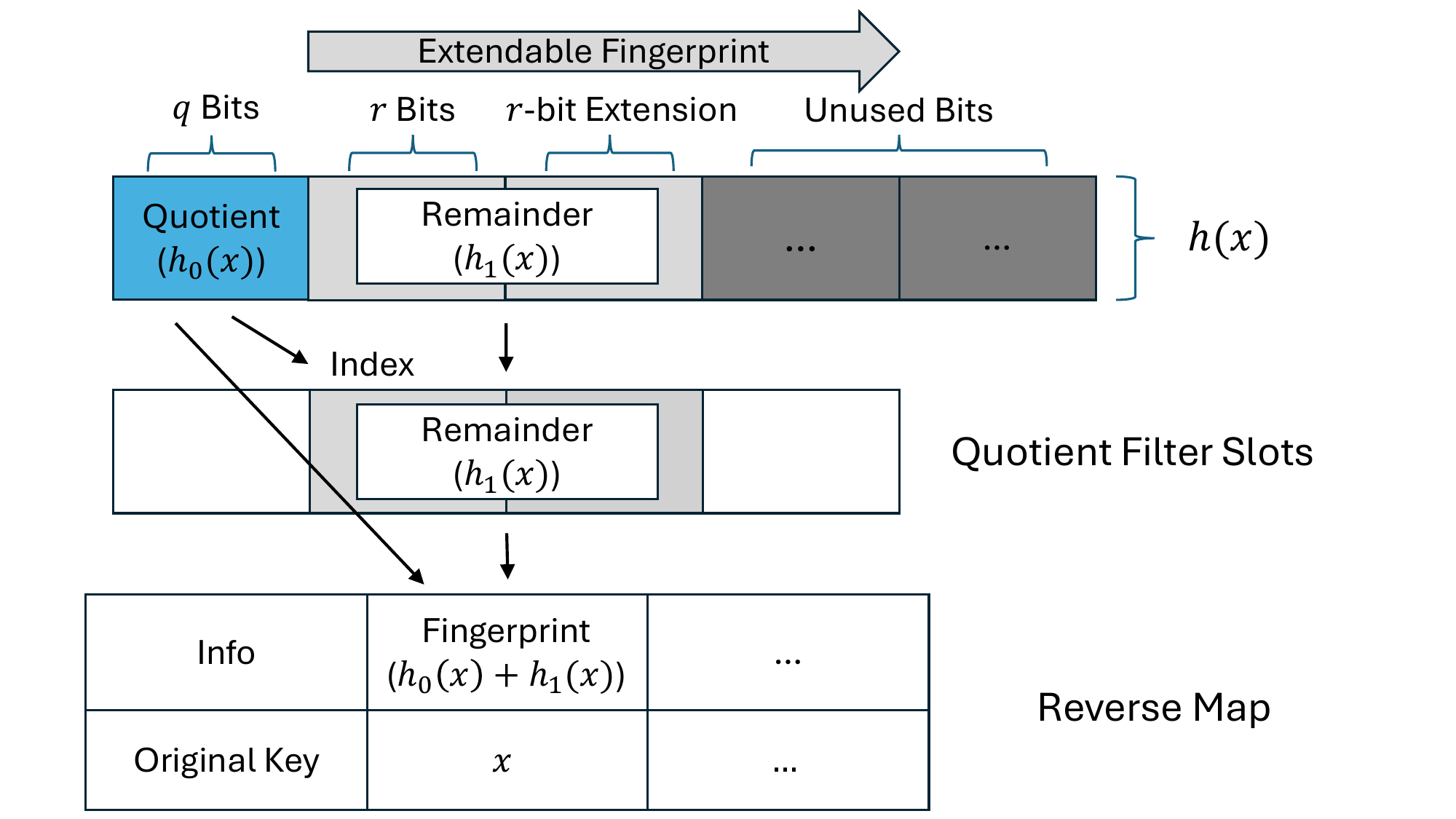}
    \caption{Example of key $x$ being inserted into an \textsc{AdaptiveQF} and its entry into the on-disk reverse map. 
    After finding the hash $h(x)$, the first $q$ bits are used to determine where the fingerprint of the hash should be stored in the filter. 
    The reverse map uses information about the fingerprint to find the original key, 
    so that the original hash can be referenced for extensions in later adaptations. 
    }
    \label{fig:aqf-ex}
\end{figure}
\fi

Additionally, we can describe the filter's space and performance:
\begin{theorem}{\cite{Wen2025}}
\begin{samepage}
    For a set of size $n$ and a target false positive probability $\epsilon$, an \textsc{AdaptiveQF} will use $(1+o(1))n\log(1/\epsilon) + O(n)$ space. If the fraction of slots in use is $\alpha$, then the cost of an insertion is $\Theta(\log n / (1-\alpha)^2)$ with high probability.
\end{samepage}
\label{thm:adaptiveperf}
\end{theorem}

The size of the filter and the runtime operations are dependent on the size
of the represented set, resulting in the quick filter operations
represented in \Cref{fig:overall-construct,fig:overall-query}.

%% file: questions.tex
\section{Gaps in modern filter evaluation}\label{sec:evaluation-gaps}

Existing research evaluates learned, adaptive, and stacked filters separately against traditional baselines 
but neglects direct comparisons between these modern approaches. 
Moreover, learned filter evaluations focus exclusively on stable workloads, 
ignoring adversarial vulnerabilities and performance under evolving distributions. 
Here, we identify key opportunities to improve filter evaluation.

\subsection{Lack of comparative analysis}
Current literature lacks extensive evaluation across modern filters: new
learned filters compare their design with other learned filters and traditional
(Bloom) filters, while adaptive filters compare their proposed design with
other adaptive filters and traditional filters. Existing literature on learned
filters does not compare them against state of the art dynamic filters such as
quotient and cuckoo filters and modern filters such as adaptive filters.
Though, stacked filters compare their performance with learned
filters~\cite{Deeds2020}, there is no comprehensive comparison of all modern
filter
types together.

Without systematic comparative analysis of these modern filter types across multiple dimensions 
including space efficiency, query performance, construction time, and robustness guarantees, 
practitioners lack the guidance needed to select appropriate filtering strategies for their specific requirements.

\subsection{Query distribution assumptions}

Current evaluations of learned filters like \textsc{Ada-BF} and \textsc{PLBF} rely on the assumption that 
query distributions will closely match training distributions.
This assumption underlies the theoretical guarantees provided by learned filters, 
as demonstrated by Mitzenmacher's~\cite{Mitzenmacher2018} theorem showing that empirical false positive rates 
remain stable when test and query sets are drawn from the same distribution. 

However, practical applications exhibit far more complex query patterns, including heavily skewed Zipfian distributions, 
dynamic workloads, and adversarial patterns where filter weaknesses are targeted.
Existing learned filter evaluations do not reflect the hostile conditions present in security-critical applications 
such as malware detection, spam filtering, or intrusion detection systems.

This evaluation gap becomes particularly concerning when considering that learned filters inherit all the vulnerabilities 
of their underlying machine learning models, including susceptibility to adversarial examples, 
model inversion attacks~\cite{Fredrikson2015}, and distribution shift exploits~\cite{Quionero-Candela2009}. 
Traditional filters provide inherent resistance to adversarial queries because their behavior depends only on 
cryptographic hash functions rather than learned patterns that can be reverse-engineered or exploited, 
and for stronger security tests new secure designs for traditional filters have been proposed~\cite{Tirmazi2025}. 
Recent work has demonstrated practical attacks on learned filters~\cite{Reviriego2021a}, 
but new designs focused on countering adaptive adversaries~\cite{Almashaqbeh2025} make assumptions 
about the number of queries in the adversarial setting.
The lack of adversarial evaluation in current literature means that the security implications of deploying 
learned filters in hostile environments remain largely unexplored, potentially creating significant vulnerabilities 
in production systems that rely on these approaches for security-critical filtering tasks.

\subsection{Missing model cost analysis}

Existing literature on learned filters exhibits a significant blind spot by
focusing exclusively on space-accuracy trade-offs while ignoring the
computational overhead introduced by model training/inference. 
These evaluations precompute the model score for each query and do \textit{not}
demonstrate the potential cost of model inference during filter queries. This
simplification misrepresents the true performance characteristics of learned
filters in practical deployments.

Comprehensive analysis reveals that model inference costs are much
higher than traditional filter operations, with the exact overhead varying
significantly based on model complexity, feature extraction requirements, and
underlying hardware. For instance, Random Forest models~\cite{Breiman2001} used
in many learned filter implementations require traversing multiple decision
trees and aggregating results, while neural network models~\cite{Scarselli2009}
involve matrix multiplications and nonlinear activations. 
Additionally, the periodic retraining required to maintain model accuracy
introduces substantial computational overheads that existing evaluations fail
to consider, making the total cost of ownership significantly higher than
traditional approaches.

While previous work did consider the query throughput of (sandwiched) learned
filters~\cite{Deeds2020}, the evaluation will benefit by extending it to the
current state of the art learned filters and adaptive filters.

\subsection{Niche use cases and dataset selection}

State-of-the-art learned filter designs are evaluated on specialized application datasets such as malware detection 
or malicious URL filtering ~\cite{Dai2020, Sato2023}, where there are naturally strong correlations between 
observable features and set membership. 
These domains represent ideal conditions for learned approaches because the features used for classification 
(file characteristics, URL structure, network patterns) have been specifically chosen for their predictive power 
in distinguishing between positive and negative examples.

However, many practical filtering applications operate on data with weaker or nonexistent feature-membership correlations. 
For example, caching systems, database query optimization, and distributed systems coordination often need to filter 
arbitrary keys or identifiers that lack meaningful extractable features. 
The emphasis on specialized datasets in current literature makes it difficult to assess how learned filters 
would perform in general-purpose applications, potentially leading to overestimation of their broad applicability. 
Furthermore, the selection bias toward datasets with strong feature correlations obscures the fundamental question of 
when learned approaches provide meaningful advantages over traditional methods.

\subsection{Dynamic workload handling}

Existing evaluations of learned filters typically assume static scenarios where both the underlying dataset and 
query patterns remain constant over time, failing to address the dynamic nature of real-world applications. 
This static evaluation methodology obscures critical questions about how learned filters degrade as workloads 
evolve and what maintenance strategies are required to sustain performance.

Real applications face continuously changing conditions including dataset growth, shifting query patterns, 
seasonal variations in access patterns, and evolving feature distributions that can rapidly obsolete trained models. 
The cost and complexity of retraining learned filters to maintain performance under these changing conditions represents a 
significant operational burden that current research largely ignores. 
Without understanding how frequently retraining is required, what triggers indicate degraded performance, 
and what graceful degradation strategies are available when retraining is delayed, it becomes impossible to assess the 
true total cost of ownership for learned filter deployments in dynamic environments.

\subsection{Theoretical versus empirical guarantees}

Traditional filters provide clear probabilistic guarantees that hold regardless of the specific data being filtered, 
enabling system designers to make precise capacity planning and performance predictions. 
In contrast, learned filters provide performance guarantees that depend on the quality of the underlying machine learning model, 
making it difficult to provide worst-case bounds essential for system design.

Mitzenmacher~\cite{Mitzenmacher2018} provides theoretical analysis and proofs describing when a learned filter 
should have better space-accuracy trade-offs compared to traditional filters. 
The proposed theoretical model provides guidance on how to size the machine learning model and the backup filters given an overall space budget. 
However, current learned filter evaluations do not explicitly validate these claims in practice, 
lacking formal guidance on how to tune configuration parameters in practical systems; 
rather, they emphasize the overall performance of the filter according to some predetermined configuration. 
While their empirical results are promising, the theoretical guarantees of learned filters remain open for validation.

%% file: app-experiments.tex
\section{Experiments and Analysis}
\subsection{Generating Zipfian queries}
\begin{figure}[t]
    \centering
    \includegraphics[scale=0.7]{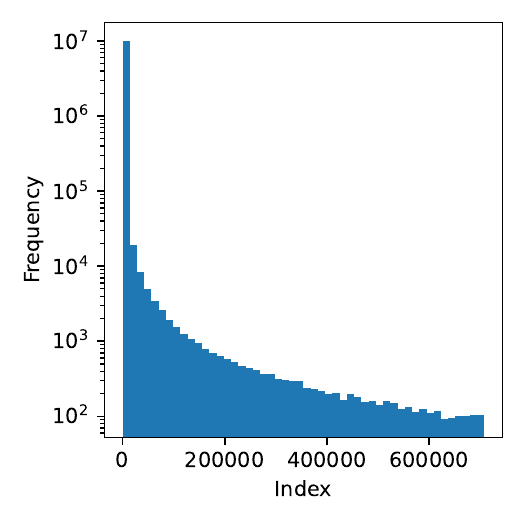}
    \caption{The query distribution for the Ember dataset prior to hashing. Following a power law,
    elements from the dataset are progressively less likely to be queried.}
    \label{fig:zipfian-unshuffled}
\end{figure}

\begin{figure}[t]
    \centering
    \includegraphics[scale=0.7]{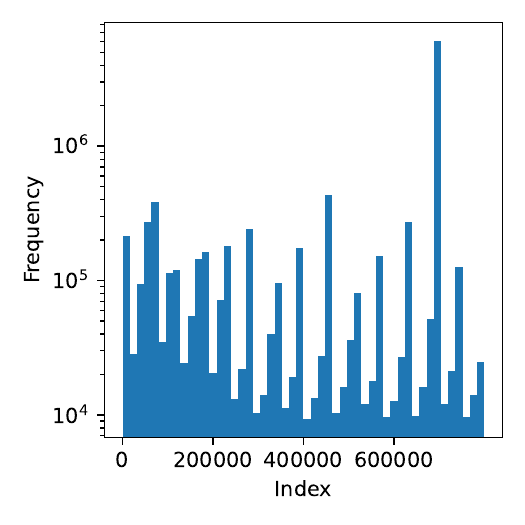}
    \caption{The query distribution for the Ember dataset after hashing. Some elements are still highly
    likely to be queried compared to others, but now the likelihood is independent of index (important since Ember is a sorted dataset).}
    \label{fig:zipfian}
\end{figure}

When generating Zipfian queries, we hash the indices of each element from a dataset
before calculating the element's likelihood of being drawn for the query workload.
In sorted datasets, hashing the indices is crucial or else either only positive or only
negative queries are more likely to be drawn.
\Cref{fig:zipfian-unshuffled} demonstrates how the Zipfian distribution on the Ember dataset appears,
while \Cref{fig:zipfian} shows an example of how hashing the indices influences
the distribution for the Ember query workload that was used in practice.


\subsection{Comparing model variants}
\begin{figure*}[h]
    \centering
    \includegraphics[width=\textwidth]{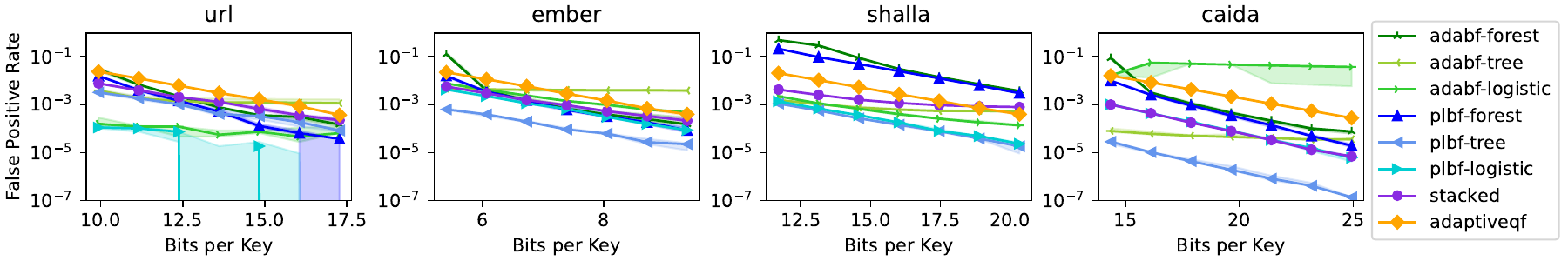}
    \caption{The FPR-space tradeoff for each dataset on a \emph{one-pass} query test.
    Learned filters perform well, with stacked filters close behind.}
    \label{fig:exp-one-pass-results}
\end{figure*}

\begin{figure*}[h]
    \centering
    \includegraphics[width=\textwidth]{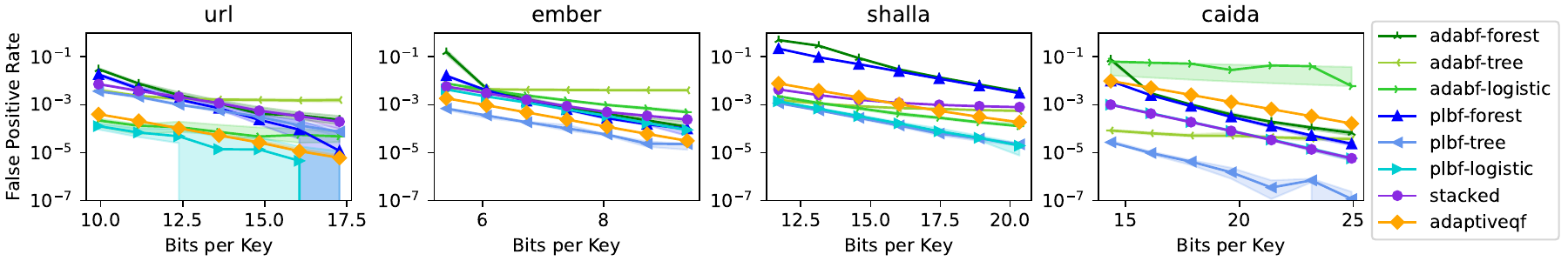}
    \caption{The FPR-space tradeoff for each dataset on 10M \emph{uniformly} distributed queries.
    Adaptive filters begin to catch up to stacked and learned filter performance.}
    \label{fig:exp-uniform-results}
\end{figure*}

\begin{figure*}[h]
    \centering
    \includegraphics[width=\textwidth]{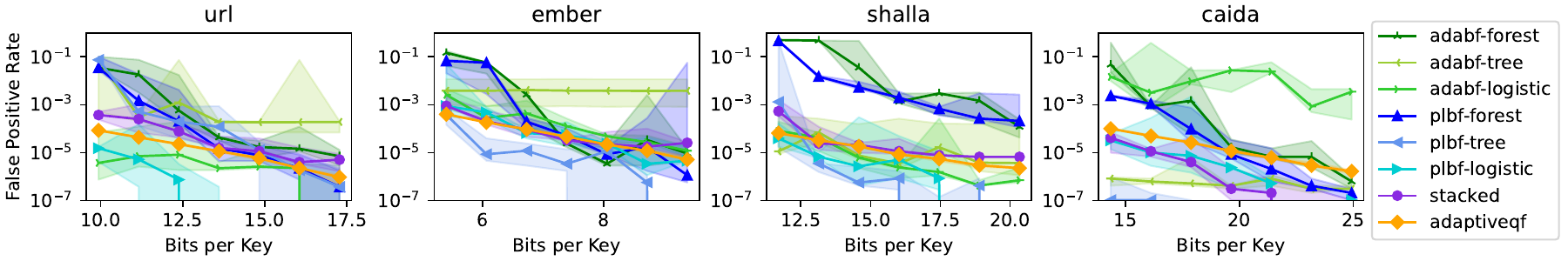}
    \caption{The FPR-space tradeoff for each dataset on 10M \emph{Zipfian} distributed queries.
    Learned filter performance becomes variable, while stacked and adaptive filters exhibit consistent and low FPRs.}
    \label{fig:exp-zipfian-results}
\end{figure*}

\begin{figure*}[h]
    \centering
    \includegraphics[width=\textwidth]{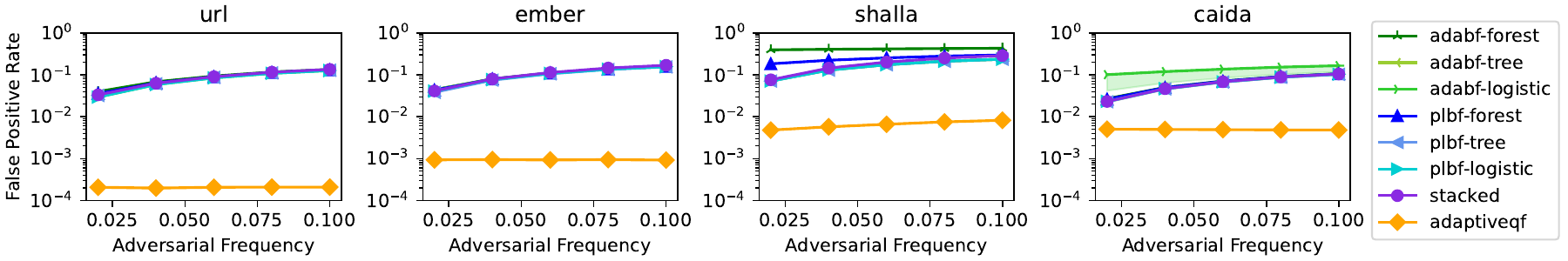}
    \caption{The effect of increasing the proportion of \emph{adversarial} queries on the FPR of fixed-size filters for each dataset.
    Adaptive filters outperform all other filter types, whose results are identical and overlap.
    }
    \label{fig:exp-adversarial-results}
\end{figure*}

\Cref{fig:exp-one-pass-results,fig:exp-uniform-results,fig:exp-zipfian-results,fig:exp-adversarial-results}
describe how all filter variants compare on the FPR experiments, now including learned filters
using random forest or logistic regression models.

\introparagraph{Random forests}
\Cref{tab:model-summary} describes that the random forest models used the largest space while still
having similar accuracies to the decision trees used in the main FPR experiments.
The new results indicate that the space usage of the random forest models make them unfit for learned filters.
Since they now have less space dedicated to backup filters while still having similar model accuracy,
learned filter variants using random forests generally have worse FPRs than decision trees.

\introparagraph{Logistic regression}
\Cref{tab:model-summary} describes that logistic regression models used the least space but demonstrated
the worst classification accuracies.
The new results indicate that this leads to inconsistent learned filter performance across different datasets.
Generally, if positive and negative key scores exhibit high overlap, learned filters are more likely
to assign positive and negative keys to the same backup filter.
Then, since the logistic regression models are small, learned filters leveraging these models
have large space budgets for backup filters.
On small datasets such as the URL dataset \cref{tab:dataset-summary}, a large space budget
for a basic backup Bloom filter will result in near-zero FPRs, resulting in learned filter variants
using logistic regression to have strong performance.
However, on other datasets their performance falters, likely due to a combination of
the backup Bloom filter's relatively low space-efficiency (compared to other traditional filter designs)
and compounding model errors.

\subsection{Dynamic experiment}
As a control of the dynamic experiment, we first perform a test where the learned filters do not retrain their models. 
We then repeat and allow the learned filters to retrain models during a churn, 
which is a common response to distribution shifts.

\begin{figure*}[h]
    \centering
    \includegraphics[width=\linewidth]{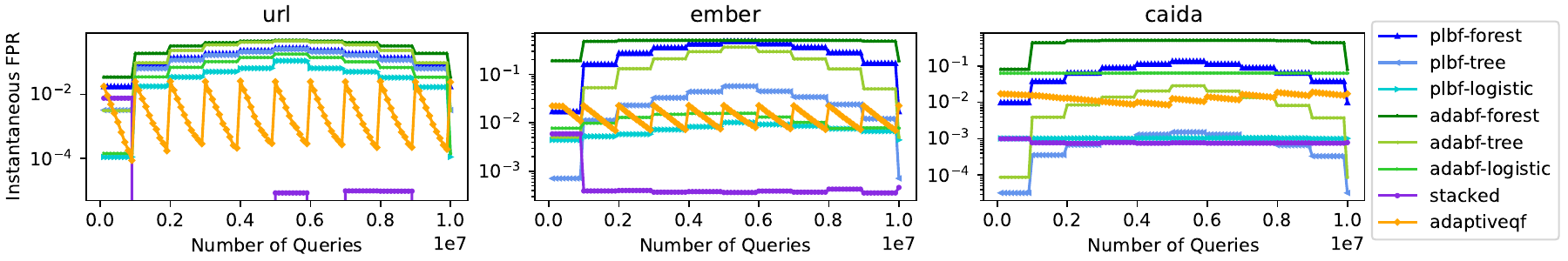}
    \caption{Instantaneous FPR of filters in response to \emph{dynamic} workload (10M uniform queries on each dataset).
    Without retraining, learned filter FPRs degrade as the positive key set changes.}
    \label{fig:dynamic}
\end{figure*}

\begin{figure*}[h]
    \centering
    \includegraphics[width=\linewidth]{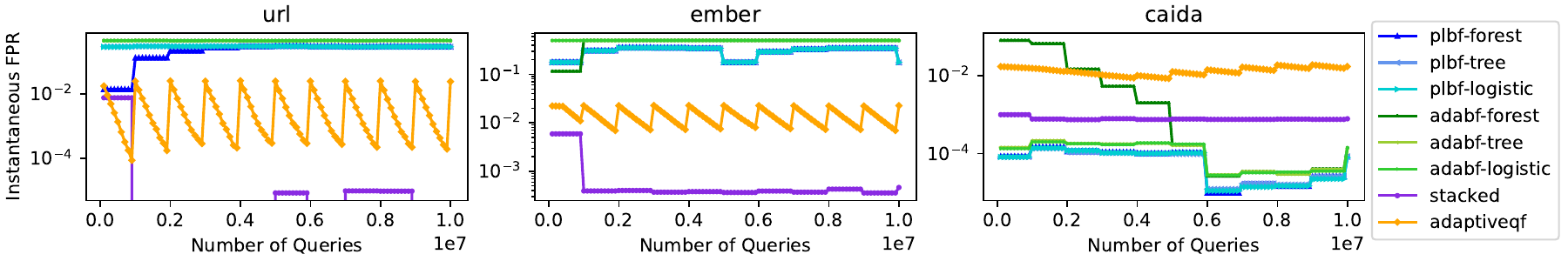}
    \caption{Instantaneous FPR of filters in response to \emph{dynamic} workload \textit{with model retraining}.
    Even with retraining, adaptive and stacked filters can still exhibit lower FPRs.}
    \label{fig:exp-dynamic-rebuild}
    \vspace{-1.5em}
\end{figure*}

\Cref{fig:dynamic} 
demonstrates that without retraining, both \textsc{Ada-BF} and
\textsc{PLBF} have performance greatly depending on how well the set they were
trained on matches the current insertion set. Between churns, their
instantaneous FPRs are fixed, occasionally at very competitive values. However,
as the dataset shifts, the model predictions across all variants become more
inaccurate, and due to lack of adaptability the filters are stuck at high FPRs,
sometimes even around $10^3\times$ worse than the \textsc{AdaptiveQF}. They
perform best only at the beginning and end of the dynamic test, when the
datasets match the original set that the models were trained on.

On all model types, \cref{fig:exp-dynamic-rebuild} still demonstrates
that retraining models during periodic churns
provides inconsistent effects on FPRs across different datasets.

\subsection{Construction and query times}
Note that for experiments involving query or construction times, key vectorization is considered as a separate step from filter operations,
since some datasets already include features and do not require any vectorization.
So, the time spent on this operation does \emph{not} contribute to the 
learned filter results for these experiments,
but in practice could serve as an additional overhead.